\documentclass[twocolumn,tighten]{aastex62}

\newcommand{\rrab}{RR{\sl ab}}
\newcommand{\rrc}{RR{\sl c}}
\newcommand{\rrd}{RR{\sl d}}

\submitjournal{ApJSS}

\shorttitle{RRLs in nearby UFDs}
\shortauthors{Vivas, Mart{\'\i}nez-V\'azquez and Walker}

\begin{document}

\title{Gaia RR Lyrae Stars in Nearby Ultra-Faint Dwarf Satellite Galaxies}

\correspondingauthor{A. K. Vivas}
\email{kvivas@ctio.noao.edu}

\author[0000-0003-4341-6172]{A. Katherina Vivas}
\affiliation{Cerro Tololo Inter-American Observatory, NSF's National Optical-Infrared Astronomy Research Laboratory, Casilla 603,
La Serena, Chile}

\author[0000-0002-9144-7726]{Clara Mart{\'\i}nez-V\'azquez}
\affiliation{Cerro Tololo Inter-American Observatory, NSF's National Optical-Infrared Astronomy Research Laboratory, Casilla 603,
La Serena, Chile}

\author[0000-0002-7123-8943]{Alistair R. Walker}
\affiliation{Cerro Tololo Inter-American Observatory, NSF's National Optical-Infrared Astronomy Research Laboratory, Casilla 603,
La Serena, Chile}

\begin{abstract}

We search for RR Lyrae stars in 27 nearby ($<100$ kpc) ultra-faint dwarf satellite galaxies using the Gaia DR2 catalog of RR Lyrae stars. Based on proper motions, magnitudes and location on the sky, we associate 47 Gaia RR Lyrae stars to 14 different satellites. Distances based on RR Lyrae stars are provided for those galaxies. We have identified RR Lyrae stars for the first time in the Tucana II dwarf galaxy, and find additional members in Ursa Major II, Coma Berenices, Hydrus I, Bootes I and Bootes III. In addition we have identified candidate extra-tidal RR Lyrae stars in six galaxies which suggest they may be undergoing tidal disruption. We found 10 galaxies have no RR Lyrae stars neither in Gaia nor in the literature. However, given the known completeness of Gaia DR2 we cannot conclude these galaxies indeed lack variable stars of this type. 
\end{abstract}


\section{Introduction} \label{sec:intro}

Ultra-Faint dwarfs (UFDs) are the most common type among the satellite galaxies of the Milky Way.  These tiny galaxies are valuable for our understanding of galaxy formation since they are the smallest dark-matter dominated systems known.
Their stars are old and very metal-poor, with little chemical enrichment. For a recent review of this type of galaxy see \citet{simon19}; following his definition of a UFD as being those having
 $M_V<-7.7$, we count 41 currently known UFDs which are satellites of the Milky Way. More than half of these were discovered in the last four years.

Observing and characterizing UFDs is challenging, and indeed not all of the new discoveries have been confirmed as UFDs. Among the list of 41 UFDs, there may be some false detections, and particularly for the fainter candidates, some systems may be globular clusters rather than galaxies. Follow-up observations to measure radial velocity dispersions and deep color-magnitude diagrams (CMD) are essential to determine the true nature of these objects. The main challenge here is the scarcity of stars, the faintest UFDs can have luminosities $<10^3 L_\sun$. In particular, the upper parts of the CMD are generally quite unpopulated, with no clear horizontal branch (HB), which makes the task of measuring an accurate distance to the UFDs very difficult. The main sequence turnoff is not generally available from the discovery (survey) photometry if the galaxy is more than $\sim50$ kpc distant, and in addition, the contamination by foreground stars and faint background galaxies may be overwhelming. Despite all these difficulties, many of the known UFDs have been investigated in detail; members have been selected via radial velocities, isochrone fitting using deep CMDs have provided ages, velocity dispersions have allowed proof of the dark matter content, and abundance analyses have permitted study of the early chemical enrichment \citep[][and references therein]{simon19}.

Observations of RR Lyrae stars (RRLs) can be complementary to the methods mentioned above. RRLs are variable stars currently in the core helium burning phase, whose progenitors were stars of $\sim 0.6-0.8  M_\sun$ \citep{catelan15}. They are old, $\gtrsim 10$ Gyrs, thus a good tracer of the type of population expected in UFDs, and are 
easy to identify because of their light variations, with amplitudes of $\sim 0.2-2.0$ magnitudes in optical bands and periods of $\sim 0.2-1.2$ d \citep{smith95}. Furthermore, and very importantly, RRLs are standard candles, which means that we can search for RRLs associated with a UFD based on their position in the sky and their mean magnitudes. If an association is found, then this allows an independent way to obtain a distance to the UFD which in principle should be more precise than an estimate from isochrone fitting, specially for the lowest luminosity systems. 
Another advantage of using RRLs for this task is that it bypasses the problem of contamination by non-members.
 Although there are field (Halo) RRLs that can reach to large distances from the Galactic center, their number density quickly declines with increasing distance \citep[e.g.][]{zinn14,medina18}. Hence, the chances of getting one or more RRLs in the location of the UFDs and at similar distance is, most of the time, negligible.

To date there has been a large effort to obtain a census of RRLs in the satellites of the Milky Way, and \citet{martinez19} presents an updated table of satellites with known RRLs, which contains 23 Galactic UFDs. Except for two cases, Carina III (Car III) and Willman I, all satellites for which a proper search has been conducted contain RRLs, although for the fainter systems the numbers can be very small. 

Here we search for RRLs in UFDs in the Gaia DR2 catalog \citep{clementini19}. Our identifications are based on position in the sky, mean magnitude and proper motions.  We have structured this paper in the following way: in \S~\ref{sec:method} we discuss the limitations of the Gaia catalog for this task and present the methodology used. We also present the sample of UFDs studied here. In \S~\ref{sec:results} we discuss our results, present the galaxies with/without RRLs, as well as the ones with extra-tidal candidates. In \S~\ref{sec:distances} we determine distances to the systems with RRLs based on Gaia photometry, and in \S~\ref{sec:conclusions} we summarize the results and discuss future work.

\section{Method} \label{sec:method}

\begin{deluxetable*}{lcccccccccccccccc}[htb!]
\rotate
\tablecolumns{17}
\tabletypesize{\scriptsize}
\tablewidth{0pc}
\tablecaption{UFD Galaxies within 100 kpc \label{tab:UFD}}
\tablehead{
Galaxy & RA & DEC & $M_V$ & Dist. & [Fe/H] & $r_h$ & $\epsilon$ & PA & $r_t$ & $N_{RR}$ & $N_{RR}$ & $N_(RR)$ & $N_{RR}$ & Ref & Ref & Ref \\
 & (deg) & (deg) & (mag) & (kpc) & (dex) & ($\arcmin$) & & (deg) & ($\arcmin$) & (lit.) & (Gaia) & (e-t) & (tot) & (param) & (p.m.) & (RRLs) \\
 (1) & (2) & (3) & (4) & (5) & (6) & (7) & (8) & (9) & (10) & (11) & (12) & (13) & (14) & (15) & (16) & (17) \\
}
\startdata
Bootes I       & 210.0200  &  14.5135  & -6.02  & 66.0  & -2.59   & 10.5  &  0.26 &   6  &  37.5 &      15      & 3     & 2     & 16  &   1 &     16 &    20,21   \\ 
Bootes III     & 209.3     &  26.8     & -5.8   & 47.0  & -2.1    & 60    &  0.5  &   90 & --    &      1       & 7     & 2     & 7   &   2 &     17 &    22      \\ 
Sagittarius II & 298.1663  & -22.065   & -5.7   & 73.1  & -2.28   & 1.7   & 0.0   & 103  & --    &      5       & 5     & 1     & 6   &   3 &      3 &    23      \\ 
Ursa Major I   & 158.7706  &  51.9479  & -5.12  & 97.0  & -2.10   & 8.13  & 0.59  &  67  & 24.0  &      7       & 6     & 0     & 7   &   1 &     16 &    24      \\ 
Hydrus I       &  37.389   & -79.3089  & -4.71  & 27.6  & -2.52   & 6.6   & 0.2   &  97  & --    &      2       & 4     & 0     & 4   &   4 &     16 &    4       \\ 
Carina II      & 114.1066  & -57.9991  & -4.5   & 37.4  & -2.44   & 8.69  & 0.34  & 170  & --    &      3       & 2     & 0     & 3   &   5 &     16 &    25      \\ 
Coma Berenices & 186.7454  &  23.9069  & -4.38  & 44.0  & -2.25   & 5.67  & 0.37  & -58  & 26.1  &      2       & 3     & 0     & 3   &   1 &     16 &    26      \\ 
UrsaMajor II   & 132.8726  &  63.1335  & -4.25  & 32.0  & -2.18   & 13.9  & 0.55  &  -76 &  59.8 &      1       & 4     & 0     & 4   &   1 &     16 &    27,28   \\ 
Triangulum II  & 33.3225  &  36.1719  & -4.2   & 28.4  & -2.6    & 2.5   & 0.3   &  73  & --    &      --      & 0     & 0     & 0   &   6,7 &   16 &    --      \\ 
Grus II        & 331.02    & -46.44    & -3.9   & 55.0  & -2.51      & 6.0   & --    &  --  & --    &      1       & 0     & 0     & 1   &   8,9,19 &   19 &    10       \\ 
Tucana II      & 343.06    & -58.57    & -3.9   & 58.0  & -2.23   & 7.2   & --    &  --  & --    &      --      & 3     & 0     & 3   &   8,10 &  16 &    --      \\ 
Reticulum II   &  53.9203  & -54.0513  & -3.88  & 30.0  & -2.46   & 5.41  & 0.56  &  69  & 19.2  &      --      & 0     & 0     & 0   &   1 &     16 &    --      \\ 
Horologium I   &  43.8813  & -54.1160  & -3.55  & 79.0  & -2.76   & 1.71  & 0.32  &   53 &  6.61 &      --      & 0     & 0     & 0   &   1 &     18 &    --      \\ 
Reticulum III  &  56.36    & -60.45    & -3.3   & 92.0  & --      & 2.4   & --    &  --  & --    &      --      & 1     & 1     & 1   &   8 &     18 &    --      \\ 
Bootes II      & 209.5141  &  12.8553  & -2.94  & 42.0  & -2.72   & 3.07  & 0.23  & -71  & 12.9  &      1       & 1     & 0     & 1   &   1 &     16 &    22      \\ 
Phoenix II     & 354.9928  & -54.4050  & -2.7   & 100   &  -2.51  &  1.5  & 0.4   &  156 &  8.14 &      1       & 1     & 0     & 1   &  9,11,12 & 18 &    9       \\
Willman I      & 162.3436  &  51.0501  & -2.53  & 38.0  & -2.11   & 2.52  & 0.47  &  74  & 16.5  &      0       & 0     & 0     & 0   &   1 &     16 &    29      \\ 
Carina III     & 114.6298  & -57.8997  & -2.4   & 27.8  & -1.97   & 3.75  & 0.55  & 150  & --    &      0       & 0     & 0     & 0   &   5 &     16 &    25      \\ 
Eridanus III   &  35.6952  & -52.2838  & -2.37  & 87.0  & -2.40   & 0.34  & 0.57  &  73  &  1.45 &      --      & 1     & 1     & 1   &   1 &     18 &    --      \\ 
Segue II       &  34.8226  &  20.1624  & -1.86  & 35.0  & -2.22   & 3.64  & 0.21  & 166  & 16.8  &      1       & 0     & 0     & 1   &   1 &     16 &    30      \\ 
Tucana V       & 354.35    & -63.27    & -1.6   & 55.0  &  -2.17   & 1.0   & 0.7   &  30  & --    &      --      & 0     & 0     & 0   &   8,19 &     19 &    --      \\ 
Horologium II  &  49.1077  & -50.0486  & -1.56  & 78.0  & -2.10   & 2.17  & 0.71  & 137  &  8.07 &      --      & 0     & 0     & 0   &   1 &     18 &    --      \\ 
Segue I        & 151.7504  &  16.0756  & -1.30  & 23.0  & -2.74   & 3.93  & 0.32  &  75  & 16.4  &      1       & 0     & 0     & 1   &   1 &     16 &    31      \\ 
Tucana III     &  359.1075 &  -59.5831 &  -1.3  & 22.9  & -2.42   & 5.1   & 0.2   &  25  & --    &      --      & 6     & 6     & 6   &   11,13 & 16 &    --      \\ 
Draco II       & 238.174   &  64.579   & -0.8   & 21.5  & -2.7    & 3.0   & 0.23  &  76  & --    &      --      & 0     & 0     & 0   &   14 &    16 &    --      \\ 
Virgo I        & 180.04    & -0.68     & -0.8   & 87.0  & --      & 1.5   & 0.44  &  51  & --    &      --      & 0     & 0     & 0   &   15 &    -- &    --      \\ 
Cetus II       &  19.47    & -17.42    &  0.0   & 30.0  & --      & 1.9   & --    &  --  & --    &      --      & 0     & 0     & 0   &   8 &     18 &    --      \\ 
\enddata
\tablecomments{Columns: (1) Galaxy name; (2) Right Ascension (J2000.0); (3) Declination (J2000); (4) Absolute Magnitude in V ; (5) Heliocentric distance; {6} metallicity; (7) half-light radius; (8) ellipticity; (9) Position angle; (10) tidal radius; (11) Number of RRLs previously known; (12) Number of RRLs found in Gaia; (13) RRLs that are extra-tidal candidates; (14) Total number of RRLs in the galaxy, including extra-tidal candidates;  (15) References for the parameters of each galaxy, including coordinates, shape parameters, distance and metallicity; (16) Reference for proper motion; (17) Reference for RRLs. }
\tablecomments{References: (1) \citet{munoz18}; (2) \citet{mcconnachie12}; (3) \citet{longeard20}; (4) \citet{koposov18}; (5) \citet{li18}; (6) \citet{carlin17}; (7) \citet{martin16};
(8) \citet{drlica15}; (9) \citet{martinez19}; (10) \citet{walker16}; (11) \citet{mutlu18}; (12) \citet{fritz19}; (13) \citet{simon17}; (14) \citet{longeard18}; (15) \citet{homma16};
(16) \citet{simon19}; (17) \citet{carlin18}; (18) \citet{pace19}; (19) \citet{simon19b}; (20) \citet{dallora06}; (21) \citet{siegel06}; (22) \citet{sesar14}; (23) \citet{joo19};
(24) \citet{garofalo13}; (25) \citet{torrealba18}; (26) \citet{musella09}; (27) \citet{dallora12}; (28) \citet{vivas16}; (29) \citet{siegel08}; (30) \citet{boettcher13}; (31) \citet{simon11}}
\end{deluxetable*} 

\begin{figure*}[htb!]
\centering
\includegraphics[width=0.32\textwidth]{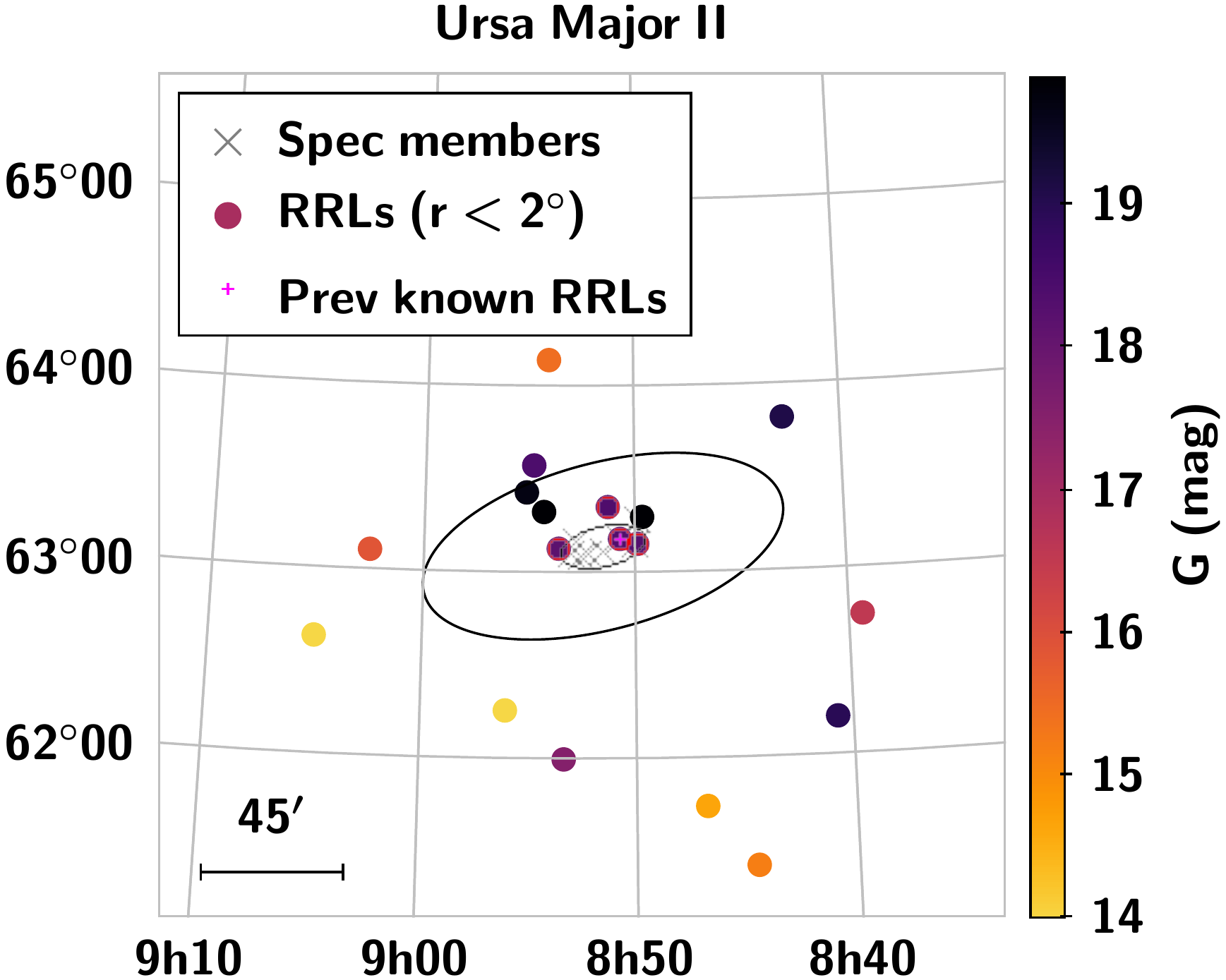}
\includegraphics[width=0.32\textwidth]{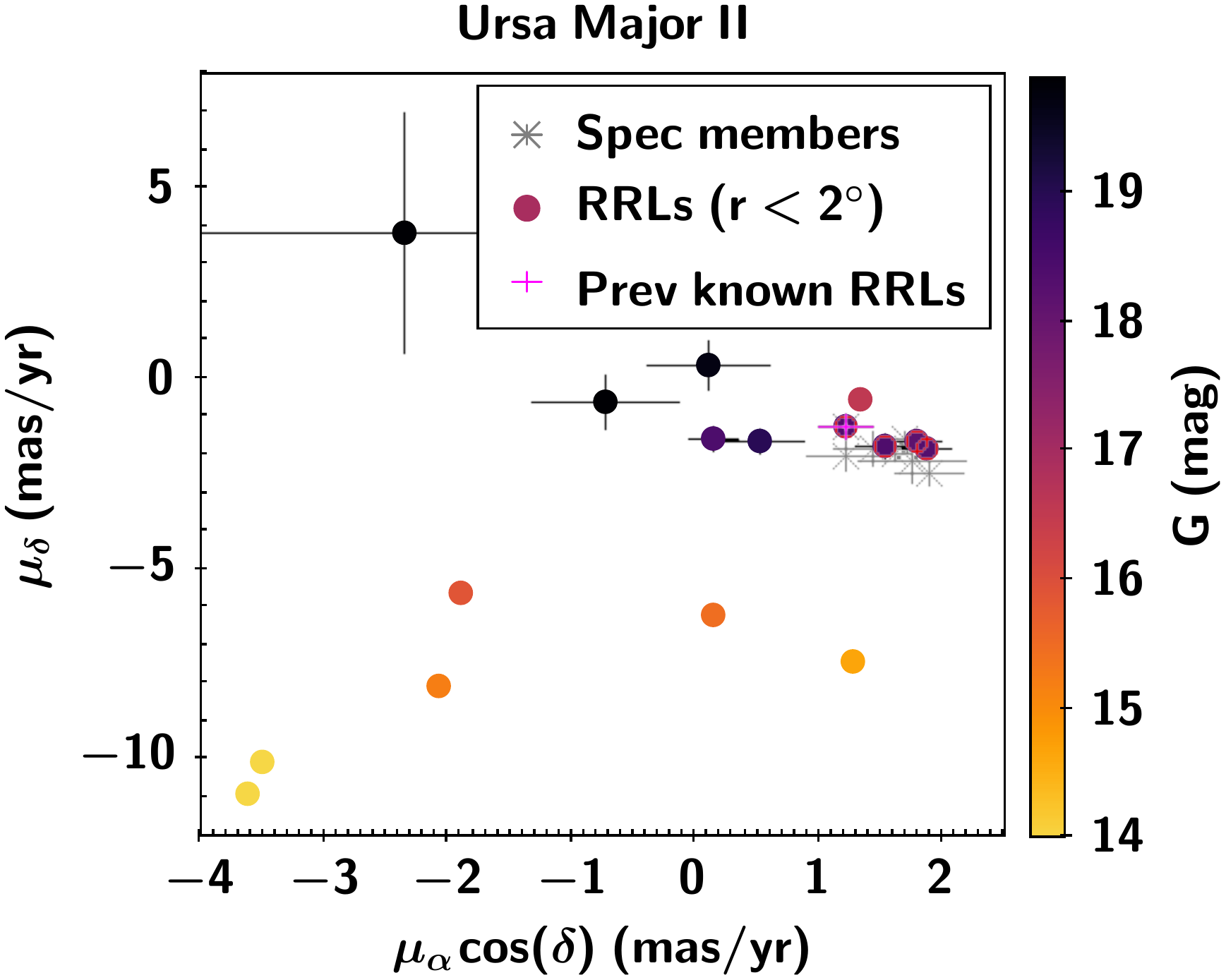}
\includegraphics[width=0.32\textwidth]{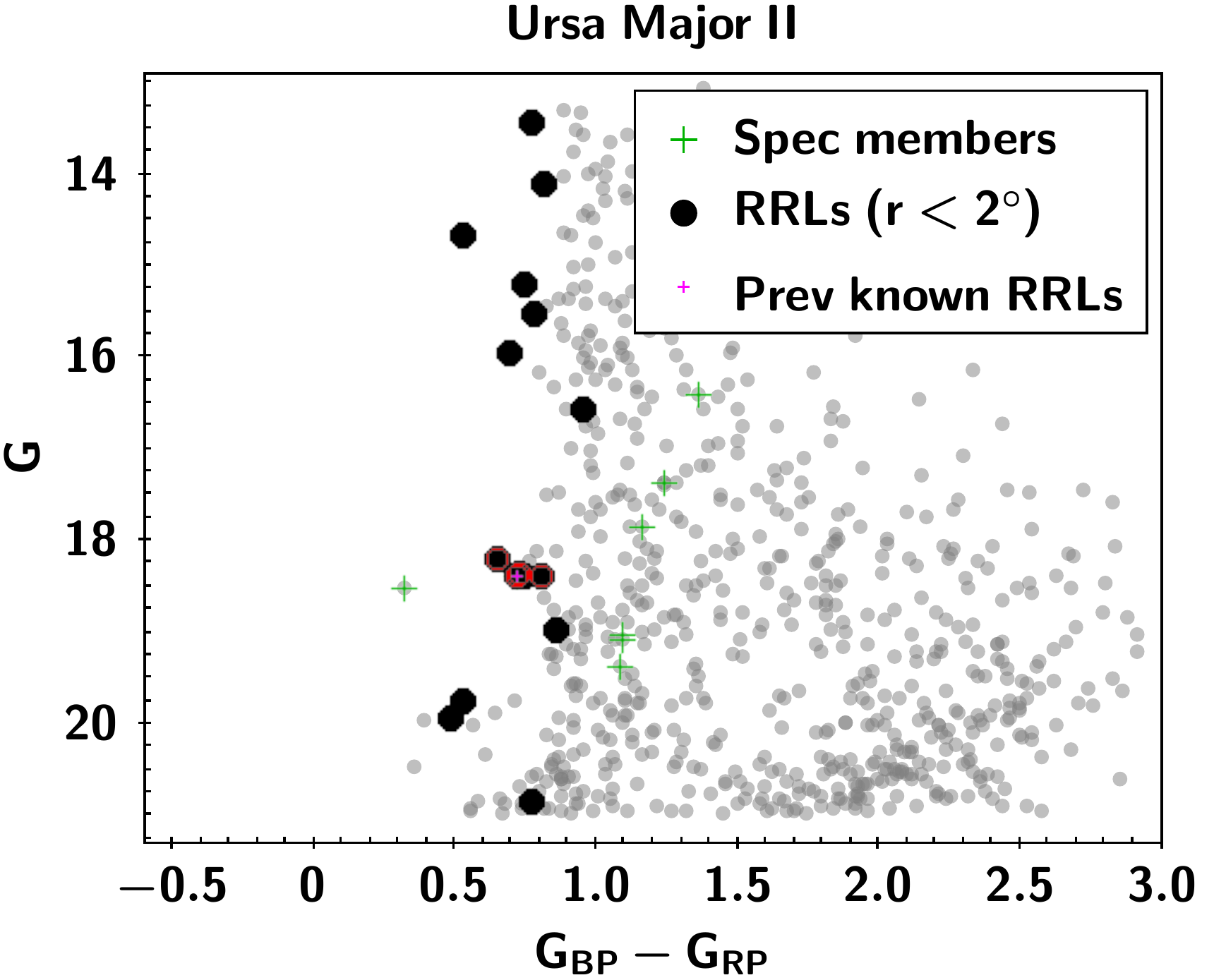}
\caption{The left and middle panels show the map in equatorial coordinates and the proper motions of RRLs within a $2\degr$ radius of UMa II (solid circles). In both panels, the color of the circles scale with the mean $G$ magnitude of the star, and spectroscopic members are shown as grey crosses. Previously known RRLs in this galaxy are marked with magenta $+$ symbols.  In the left panel, the inner and outer ellipses indicate the half-light radius and the tidal radius ($r_t$), respectively (see Table~\ref{tab:UFD}). The right panel is a Gaia CMD (grey background) of UMa II stars within $1\, r_h$ from its center. The RRLs which are identified as members of the galaxy are enclosed by open red circles in all panels. }
\label{fig:UMaII}
\end{figure*}

\citet{clementini19} present a catalog\footnote{Table \texttt{gaiadr2.vari\_rrlyrae}}  of 140,784 RRLs from Gaia DR2. Although they acknowledge some of these RRLs are associated with UFDs, they do not show the specific findings for these galaxies. RRLs in Gaia DR2 have mean magnitudes between $9\lesssim G \lesssim21$. The catalog is known to be incomplete, with the completeness function heavily depending on position in the sky since some parts of the sky have been observed more frequently than others, and consequently the sampling of the light curves in some cases is poor enough to not allow recognition of a star as a RRL. \citet{clementini19} cite an average completeness of $60\%$.   The advantages of the Gaia catalog of RRLs is that it is all-sky, and goes deeper than any other previous large survey available in the literature. 

Our method for searching for RRLs in UFDs is as follows. We compiled a list of all UFD galaxies at distances less than 100 kpc, since more distant galaxies would have RRLs beyond the Gaia DR2 limits \footnote {Distant UFDs not included here are Hercules, Leo IV, Leo V, Hydra II, Eridanus II, Pictor I, Grus I, Columba I, Indus II, Canes Venatici II, Pisces II, Pegasus III, Aquarius II, and Bootes IV.}. The list of selected UFDs (27 galaxies) is shown in Table~\ref{tab:UFD}. The list contains all UFDs known to the time of writing this paper. Structural parameters such as size, position angle (PA), and ellipticity for many of these galaxies were recently updated by \citet{munoz18}. For those not in their list, we assumed the structural parameters in the discovery papers or in photometric follow-up papers. The number of known RRLs in UFDs was taken from the recent compilation by \citet{martinez19}. Table~\ref{tab:UFD} contains in column (12) the number of RRLs found in Gaia for each galaxy. Some of these were already known, but others are new identifications as members of UFDs.  Within our limit of distance $<100$ kpc we considered all galaxies whether or not they have been searched before for RRLs, since Gaia may contain new members, for example, in the outermost regions of a galaxy which may not have been covered by previous observations.

In addition we compiled proper motion information for confirmed members of each galaxy. These data were taken mostly from \citet{simon18} and \citet{pace19}. For the latter, no radial velocity measurements are available, so the selected stars are only proper motion-selected members. However, we only considered stars with high probability ($>0.8$) of being a proper motion member. For Bootes III (Boo III) and Sagittarius II (Sgr II), members were taken from \citet{carlin18} and \citet{longeard20}, respectively. Spectroscopic members of Grus II and Tucana V (Tuc V) come from the recent work by \citet{simon19b}.

We then selected from the Gaia DR2 catalog of RRLs all stars within a circle of radius $1\degr$. The search area was increased to a $2\degr$ radius for the larger Bootes I (Boo I), Boo III, Ursa Major II (UMa II), and for the disrupting galaxy Tucana III (Tuc III). Most galaxies have half-light radii ($r_h$) much smaller than this size (see column (7) in Table~\ref{tab:UFD}). The left panel in Figure~\ref{fig:UMaII} shows the case for UMa II. Within a $2\degr$ radius there are 18 RRLs in the Gaia catalog, several of them inside the tidal radius of the galaxy. The color of the RRLs in this plot scales with the mean magnitude of the star. We also identified previously known RRLs in the galaxy with magenta $+$ symbols. The source of the previously known RRLs are in column (17) of Table~\ref{tab:UFD}. UMa II has one, which was also identified in Gaia. We intentionally explored a very large area in order to search for distant possible members which may be candidates for being debris material of galaxies under tidal disruption. Several UFDs are believed to have suffered tidal disruption, with the most clear case being the Tuc III galaxy \citep{drlica15,shipp18,li18b}, which shows clear tidal tails to each side of the dwarf. The Hercules UFD is also believed to be suffering tidal disruption and indeed, extra-tidal RRLs have been identified in this galaxy \citep{garling18}.

We compared the proper motion of the selected RRLs with those for known members in each galaxy. In the middle panel of Figure~\ref{fig:UMaII}, the proper motions of radial velocity members of UMa II are shown as grey crosses. In general, if a galaxy had previously known RRLs, we marked them with magenta $+$ symbols. This was done even if the RRLs do not exist in the Gaia catalog of RRLs but only in the main DR2 catalog. The proper motion of the RRLs within the search area for each galaxy are plotted with their error bars, with colors that scale with their mean magnitude. Notice that for bright stars sometimes the error bars are smaller than the symbol size. It is clear that only a few RRLs share the same proper motion as the UFD galaxy. All other RRLs in the line of sight are most likely Halo field stars.

Additional constraints are made based on the mean $G$ magnitude of the RRLs and the location on the CMD. We expect all RRLs in a galaxy to have similar magnitude since they all lie in the HB of the galaxy \citep[and there is no dependence of magnitude with period in the $G$ band][]{muraveva18}. The right panel in Figure~\ref{fig:UMaII} shows the Gaia CMD of UMa II. Because UFDs have very few stars, in order to be able to minimize field contamination and be able to distinguish any feature of the galaxy, we limit the CMDs to stars  within $1\, r_h$ of each galaxy (or $2\,r_h$ for the smallest galaxies with $r_h<3\arcmin$ since the number of stars in the CMD is otherwise too low). If there were previously known RRLs in that galaxy (magenta $+$'s), we know the magnitude they should have. 
In the case of UMa II, there are three additional RRLs with the same magnitude ($G\sim 18.3$) as the previously known one. When no previously known RRLs exist, there are estimates of the distance to each galaxy from the literature (column 5 in Table~\ref{tab:UFD}) which provides an approximate magnitude for the HB. The CMDs were useful also to check the color of the RRLs. It is known that there are some misclassifications in the Gaia RRLs catalog \citep{clementini19}. Indeed, some of our CMDs show some very red RRLs, which are likely misclassifications.

The combination of the CMDs and proper motion plots similar to those in Figure~\ref{fig:UMaII}  were adequate to identify RRL stars in the UFDs. In the case of UMa II, the 4 RRLs with the same magnitude are also the ones sharing the same proper motion as the galaxy. Our selected RRLs are shown encircled in red in all panels in Figure~\ref{fig:UMaII}.  In UMa II, all of the identified RRLs lie within the limits of the galaxy. There are however some interesting cases in which extra tidal stars may have been detected.

As a final step, we checked the individual Gaia lightcurves of the selected stars, available from the Gaia DR2 site, as well as image stamps in Aladin.  Two (out of an initial selection of 49 RRLs belonging to UFDs) turned up to have very dubious phased lightcurves and, in addition, are listed as likely misclassifications in Appendix C of \citet{clementini19}.

Table~\ref{tab:UFD} summarizes our results in columns 11-14. These columns contain the number of previously know RRLs, the RRLs found in Gaia with the method above, the number of candidate  extra-tidal RRLs, and the total number of RRLs (adding both Gaia and the literature) including the extra-tidal candidates.

\section{Results} \label{sec:results}

 With the methodology described above we were able to identify 47 RRLs in 14 UFDs. Of them, 24 RRLs are new identifications as UFD members. The other stars were previously known. We note that Gaia did not identify all known RRLs in UFDs.  In the following subsections we discuss in detail our findings for each galaxy.

\longrotatetable
\begin{deluxetable*}{lccccccccccccccc}
\tabcolsep=0.1cm
\movetabledown=1cm
\tablecolumns{16}
\tabletypesize{\scriptsize}
\tablewidth{0pc}
\tablecaption{RRLs found in Gaia DR2 in UFD Galaxies within 100 kpc \label{tab:RRL}}
\tablehead{
ID & Source ID & RA & Dec & $\mu_{\alpha}\cos{\delta}$ & $\mu_{\delta}$ & Period & $\langle G \rangle$ & Amp G & Classif. &  D$_{Host}$ & E($B-V$) &  D$_{\odot}$ &  $\sigma_{D_{\odot}}$ & Previous ID & Previous Ref.\\
% & (deg) & (deg) & (mas/yr) & (mas/yr) & (days) & (mag) & (mag) &  & (arcmin) & & (mag) & (mag) & (mag) & (mag) & (mag) & (kpc) & (kpc) & & \\
 (1) & (2) & (3) & (4) & (5) & (6) & (7) & (8) & (9) & (10) & (11) & (12) & (13) & (14) & (15) & (16) \\
}
\startdata
  Boo I-V8 &  1230833517027065728 &  209.998692 &  14.459412 & -0.93 &  -0.86 &  0.417774 &          19.43 &            0.43 &                RRc &             3.5 &     0.017 &     66 &           7 &          V8 &    1, 2 \\
  Boo I-V11\tablenotemark{*} &  1230741020611316224 &  209.518254 &  14.222002 & -0.26 &  -0.73 &  0.661748 &          19.52 &            0.82 &                RRab &            34.0 &     0.019 &     69 &           7 &         V11 &       1 \\
  Boo I-V16\tablenotemark{*} &  1230914571649196544 &  208.645228 &  14.204217 & -0.42 &   0.27 &  0.347504 &          19.16 &            0.46 &                 RRc &            82.0 &     0.022 &     58 &           6 &          -- &      -- \\
  Boo II-V1 &  3727826519650056576 &  209.529333 &  12.856341 & -2.74 &  -0.49 &  0.663511 &          18.32 &            0.61 &                RRab &             0.9 &     0.030 &     40 &           4 &          -- &       3 \\
 Boo III-V1 &  1258556500130302080 &  210.143865 &  25.931296 & -1.34 &  -0.98 &  0.633280 &          18.75 &            0.94 &                RRab &            69.1 &     0.015 &     45 &           4 &          -- &       3 \\
 Boo III-V2 &  1451041850411971840 &  209.312082 &  27.120035 & -1.18 &  -1.03 &  0.763933 &          18.74 &            0.26 &                RRab &            19.2 &     0.021 &     45 &           4 &          -- &      -- \\
 Boo III-V3 &  1450796178282259072 &  209.222201 &  26.465477 & -1.32 &  -0.62 &  0.591339 &          18.80 &            1.28 &                RRab &            20.5 &     0.017 &     46 &           4 &          -- &      -- \\
 Boo III-V4 &  1450755118394910720 &  209.529019 &  26.471237 & -1.23 &  -0.13 &  0.616001 &          18.82 &            0.35 &                RRab &            23.2 &     0.018 &     46 &           4 &          -- &      -- \\
 Boo III-V5 &  1450750170592551040 &  209.529158 &  26.385552 & -2.09 &  -0.47 &  0.371664 &          18.84 &            0.50 &                 RRc &            27.7 &     0.016 &     47 &           4 &          -- &      -- \\
  Boo III-V6\tablenotemark{*} &  1258709504045092608 &  210.343830 &  26.650132 & -1.11 &  -0.27 &  0.405316 &          18.86 &            0.34 &                 RRc &            56.7 &     0.016 &     47 &           4 &          -- &      -- \\
  Boo III-V7\tablenotemark{*} &  1450391626723098496 &  208.384631 &  25.950000 & -1.88 &  -1.25 &  0.344303 &          18.95 &            0.58 &                 RRc &            70.9 &     0.014 &     49 &           4 &          -- &      -- \\
  Car II-V2 &  5293954286501519616 &  114.191046 & -57.865022 &  2.27 &   0.17 &  0.407954 &          18.56 &            0.43 &                 RRc &             8.5 &     0.192 &     35 &           3 &   CarII\_V2 &       4 \\
  Car II-V3 &  5293940924860019584 &  113.787986 & -57.954102 &  2.21 &   0.04 &  0.705140 &          18.46 &            0.54 &                RRab &            10.5 &     0.177 &     34 &           3 &   CarII\_V3 &       4 \\
  ComBer-V2 &  3959868759245750400 &  186.711937 &  23.933415 &  0.94 &  -2.14 &  0.310087 &          18.61 &            0.31 &                 RRc &             2.4 &     0.018 &     43 &           4 &          V2 &       5 \\
  ComBer-V1 &  3959870167995015936 &  186.889598 &  23.915354 &  1.18 &  -1.95 &  0.669899 &          18.34 &            0.92 &                RRab &             7.9 &     0.018 &     38 &           4 &          V1 &       5 \\
  ComBer-V4 &  3959816395004385152 &  187.038475 &  23.657540 &  0.37 &  -1.73 &  0.669770 &          18.37 &            0.79 &                RRab &            22.0 &     0.018 &     39 &           4 &          -- &      -- \\
 Eri III-V1\tablenotemark{*} &  4744205275840865408 &   34.751750 & -52.778135 &  2.19 &   0.72 &  0.337076 &          20.40 &            0.32 &                 RRc &            45.4 &     0.036 &     98 &           9 &          -- &      -- \\
 Hyd I-V1 &  4632586467457695872 &   37.433060 & -79.277316 &  3.96 &  -1.32 &  0.670820 &          17.95 &            0.69 &                RRab &             2.0 &     0.091 &     30 &           3 &       6316\tablenotemark{**} &       6 \\
 Hyd I-V2 &  4632533789682809344 &   38.126669 & -79.457534 &  3.55 &  -1.39 &  0.724643 &          17.89 &            0.46 &                RRab &            12.1 &     0.097 &     29 &           3 &       6325\tablenotemark{**} &       6 \\
 Hyd I-V3 &  4632605704614645632 &   36.987668 & -79.052077 &  3.36 &  -1.58 &  0.395873 &          17.72 &            0.66 &                RRab &            16.1 &     0.095 &     27 &           3 &          -- &      -- \\
 Hyd I-V4 &  4632155248445477888 &   37.349951 & -79.661236 &  4.13 &  -1.17 &  0.631974 &          17.97 &            0.62 &                RRab &            21.1 &     0.104 &     30 &           3 &          -- &      -- \\
  Phe II-V1 &  6497787714959165440 &  354.929687 & -54.422860 & -0.24 &  -1.90 &  0.608239 &          20.25 &            1.23 &                RRab &             2.6 &     0.013 &     99 &          10 &          V1 &      7 \\
  Ret III-V1\tablenotemark{*} &  4681004144885429248 &   57.551475 & -59.877889 & -1.59 &  -1.85 &  0.621006 &          20.13 &            0.32 &                RRab &            49.4 &     0.028 &     87 &           8 &          -- &      -- \\  
  Sgr II-V2 &  6864422757758521984 &  298.236471 & -22.069005 &  0.97 &  -1.76 &  0.406520 &          19.63 &            0.41 &                 RRc &             3.9 &     0.107 &     62 &           6 &          V2 &       8 \\
  Sgr II-V3 &  6864422993976659968 &  298.184989 & -22.051337 &  1.69 &  -1.12 &  0.665683 &          19.63 &            0.64 &                RRab &             1.3 &     0.110 &     62 &           6 &          V3 &       8 \\
  Sgr II-V4\tablenotemark{*} &  6865195302117134336 &  298.054710 & -21.716383 & -1.39 &  -1.19 &  0.540729 &          19.46 &            0.78 &                RRab &            21.8 &     0.103 &     58 &           5 &          V4 &       8 \\
  Sgr II-V5 &  6864048408410304896 &  298.158548 & -22.058483 &  0.87 &  -0.73 &  0.307847 &          19.77 &            0.53 &                 RRc &             0.6 &     0.111 &     66 &           6 &          V5 &       8 \\
  Sgr II-V6 &  6864423994704275200 &  298.149510 & -22.033303 &  0.60 &  -0.76 &  0.318561 &          19.76 &            0.55 &                 RRc &             2.1 &     0.112 &     65 &           6 &          V6 &       8 \\
  Tuc II-V1 &  6503773559340262144 &  342.901932 & -58.504744 &  0.66 &  -1.51 &  0.534074 &          19.47 &            0.66 &                RRab &             4.6 &     0.020 &     64 &           6 &          -- &      -- \\
  Tuc II-V2 &  6503773902937660160 &  342.887268 & -58.475085 &  0.88 &  -1.36 &  0.303942 &          19.05 &            0.27 &                 RRc &             6.4 &     0.019 &     52 &           5 &          -- &      -- \\
  Tuc II-V3 &  6491895638304405632 &  343.778317 & -58.063979 &  0.58 &  -0.97 &  0.318948 &          19.08 &            0.51 &                 RRc &            39.4 &     0.017 &     53 &           5 &          -- &      -- \\
 Tuc III-V1\tablenotemark{*} &  4918034941751885696 &    1.072293 & -59.376886 & -0.19 &  -1.47 &  0.572852 &          17.45 &            0.77 &                RRab &            60.1 &     0.012 &     26 &           3 &          -- &      -- \\
 Tuc III-V2\tablenotemark{*} &  4905943165704407808 &    1.674563 & -60.158202 &  0.86 &  -2.36 &  0.684696 &          17.15 &            0.39 &                RRab &            83.1 &     0.011 &     23 &           2 &          -- &      -- \\
 Tuc III-V3\tablenotemark{*} &  4918327235750088576 &    1.738337 & -58.663973 &  0.20 &  -1.27 &  0.559389 &          17.52 &            1.00 &                RRab &            97.5 &     0.011 &     27 &           3 &          -- &      -- \\
 Tuc III-V4\tablenotemark{*} &  6494865556649868928 &  358.457911 & -57.920035 &  0.89 &  -1.69 &  0.311229 &          17.74 &            0.41 &                 RRc &           103.1 &     0.013 &     30 &           3 &          -- &      -- \\
 Tuc III-V5\tablenotemark{*} &  6494902222786232192 &  359.594316 & -57.877537 &  0.97 &  -1.70 &  0.526733 &          17.39 &            1.04 &                RRab &           104.3 &     0.014 &     25 &           2 &          -- &      -- \\
 Tuc III-V6\tablenotemark{*} &  6489171190225163776 &  355.498432 & -59.677556 & -0.26 &  -1.51 &  0.599675 &          17.12 &            0.72 &                RRab &           110.8 &     0.016 &     22 &           2 &          -- &      -- \\
  UMa I-V1 &   847520295882556928 &  158.746852 &  51.951965 &  1.69 &  -2.06 &  0.569578 &          20.48 &            1.13 &                RRab &             0.9 &     0.018 &     99 &           9 &          V1 &       9 \\
  UMa I-V2 &   847519849205957760 &  158.773319 &  51.927926 & -1.94 &  -0.93 &  0.584289 &          20.51 &            1.04 &                RRab &             1.2 &     0.019 &    101 &           9 &          V2 &       9 \\
  UMa I-V3 &   847706663103625984 &  158.628336 &  51.941272 & -1.31 &  -0.45 &  0.643148 &          20.38 &            1.13 &                RRab &             5.3 &     0.019 &     95 &           8 &          V3 &       9 \\
  UMa I-V4 &   847707728255299712 &  158.577970 &  51.974729 & -2.11 &   0.51 &  0.745356 &          20.22 &            0.81 &                RRab &             7.3 &     0.018 &     88 &           8 &          V4 &       9 \\
  UMa I-V5 &   849022228766244992 &  158.906709 &  52.043444 & -0.92 &   0.04 &  0.599648 &          20.46 &            0.86 &                RRab &             7.6 &     0.017 &     99 &           9 &          V5 &       9 \\
  UMa I-V6 &   847700718868881792 &  158.280031 &  51.834913 & -2.25 &  -0.48 &  0.39673 &          20.49 &            0.44 &                RRab &            19.4 &     0.019 &    100 &           9 &          V6 &       9 \\
  UMa II-V1 &  1043841876592990208 &  132.656424 &  63.169076 &  1.23 &  -1.27 &  0.565120 &          18.33 &            1.14 &                RRab &             6.2 &     0.105 &     34 &           3 &          V1 &       10,11 \\
  UMa II-V2 &  1043842254550107648 &  132.448540 &  63.141615 &  1.87 &  -1.89 &  0.568795 &          18.31 &            0.77 &                RRab &            11.5 &     0.112 &     33 &           3 &          -- &      -- \\
  UMa II-V3 &  1043938427457808384 &  132.811179 &  63.342705 &  1.53 &  -1.80 &  0.402186 &          18.37 &            0.44 &                 RRc &            12.7 &     0.118 &     34 &           3 &          -- &      -- \\
  UMa II-V4 &  1043873010812474752 &  133.382084 &  63.113076 &  1.79 &  -1.70 &  0.411628 &          18.17 &            0.39 &                RRc &            13.9 &     0.075 &     32 &           3 &          -- &      -- \\
\enddata
\tablenotetext{*}{Extra-tidal candidate}
\tablenotetext{**}{The complete name of these stars is preceded by the prefix ``OGLE-SMC-RRLYR-''.}
\tablecomments{Columns: (1) RRL identification in this work preceded by the name of the host galaxy;
(2) Unique source identifier in Gaia; (3) Right Ascension in degrees (J2015.5) ; (4) Declination (J2015.5) in degrees; 
(5) Proper motion in RA direction (mas/yr); (6) Proper motion in Dec direction (mas/yr); (7) Period (days); 
(8) Intensity-averaged magnitude in the G band; (9) Peak-to-peak amplitude of the G band light curve; 
(10) Best RR Lyrae classification estimate; (11) Distance to the host galaxy (in arcmin);
(12) Interstellar dust reddening; (13) Distance to the Sun (kpc); (14) Uncertainty of the distance to the Sun (kpc);
(15) Previous identifications; (16) References to the previous identifications.}
\tablecomments{References: (1) \citet{siegel06}; (2) \citet{dallora06};  (3) \citet{sesar14}; (4) \citet{torrealba18}; (5) \citet{musella09}; (6) \citet{koposov18}; (7) \citet{martinez19}; (8) \citet{joo19}; (9) \citet{garofalo13}; (10) \citet{dallora12}; (11) \citet{vivas16} }
\end{deluxetable*} 
\endlongrotatetable

\subsection{UFD Galaxies with Gaia RRLs}

Table~\ref{tab:RRL} contains the IDs, coordinates, proper motions, light curve properties, distance to the center of the host galaxy, and heliocentric distance (\S~\ref{sec:distances}) of all the Gaia RRLs found in UFDs. Coordinates, proper motions, periods, intensity-averaged $G$ magnitudes, amplitudes in the $G$ band, and types are all taken from the Gaia DR2 catalog (tables \texttt{gaiadr2.gaia\_source} and \texttt{gaiadr2.vari$\_$rrlyrae}). The only exceptions are the types for stars Boo I V8 and UMa I V4 which were changed from {\rrab } to {\rrc } after inspecting their lightcurves, as well as the period of UMa I V6 for which we adopted the period given in \citet{garofalo13} because it produces a better phased light curve of the Gaia data. The new classification for Boo I V8 agrees with the one previously given by \citet{siegel06}.  Gaia $G$-band phased lightcurves are provided as online-only figure set~\ref{fig:lc} in the Appendix section (\S~\ref{sec:appendix}).

Gaia recovered known RRLs in several UFDs. These stars are included in Table~\ref{tab:RRL} with their original IDs. Notice however that in some cases, Gaia did not recover all of the known RRLs in a particular galaxy. The most striking case is Bootes I (Boo I) which is known to contain a large population of RRLs \citep[15,][]{siegel06}. We only found two of them in Gaia. This seems to indicate that the completeness of Gaia in the part of the sky where Boo I is located is particularly low. For other cases, the recovery rate is not so low. In Sgr II, for example, Gaia recovers five out of the six known RRLs, 6/7 in Ursa Major I (UMa I), and 1/1 in Bootes II (Boo II) and Phoenix II (Phe II).  More interestingly, we were able to associate new RRLs from the Gaia catalog to some UFDs. In some cases these are the first RRLs detected in that galaxy. In others, Gaia found additional RRL members to the ones already known.

We discuss the UFD galaxies with Gaia RRLs in three groups: {\sl (i)} four galaxies with new RRLs members, all of them located within their tidal radius, {\sl (ii)} six galaxies with new members including extra-tidal candidates, and {\sl (iii)} four galaxies for which we recovered known RRLs in Gaia but no new members can be associated with them.

\subsubsection{UFD galaxies with new RRLs members within their tidal radius}

\paragraph{Ursa Major II (Figure~\ref{fig:UMaII})}

UMa~II is one of the faintest among the SDSS UFDs. Its elongated shape may suggest it is undergoing tidal disruption. \citet{dallora12} found one RRL in UMa~II but with doubts about its period. The period was revised in \citet{vivas16} using data from the Catalina Real Time Transient Survey (CRTS), finding that it was indeed different to the one suggested by \citet{dallora12}. Gaia finds this star with the same period as \citeauthor{vivas16}  Figure~\ref{fig:UMaII} shows a total of four RRLs of about the same magnitude (same color in the plot) in agreement with the proper motion of the spectroscopic members of UMa~II. These stars have magnitudes between $18.18 < G < 18.37$. In the sky (right panel), those four stars (encircled in red) are all relatively close to the center of UMa~II, at distances between $5\arcmin$ to $14\arcmin$. Since the tidal radius of this galaxy has been estimated to be $59\farcm 8$ \citep{munoz18}, we conclude these stars must be members. Thus, UMa~II has a population of four RRLs, of which three are {\rrab } and one is \rrc. The mean period of the three {\rrab } is 0.515d, which is short for RRLs in UFDs. Based on the mean period, UMa~II would be classified as an Oosterhoff (Oo) I system \citep{oosterhoff39}.

\begin{figure*}[htb!]
\centering
\includegraphics[width=0.32\textwidth]{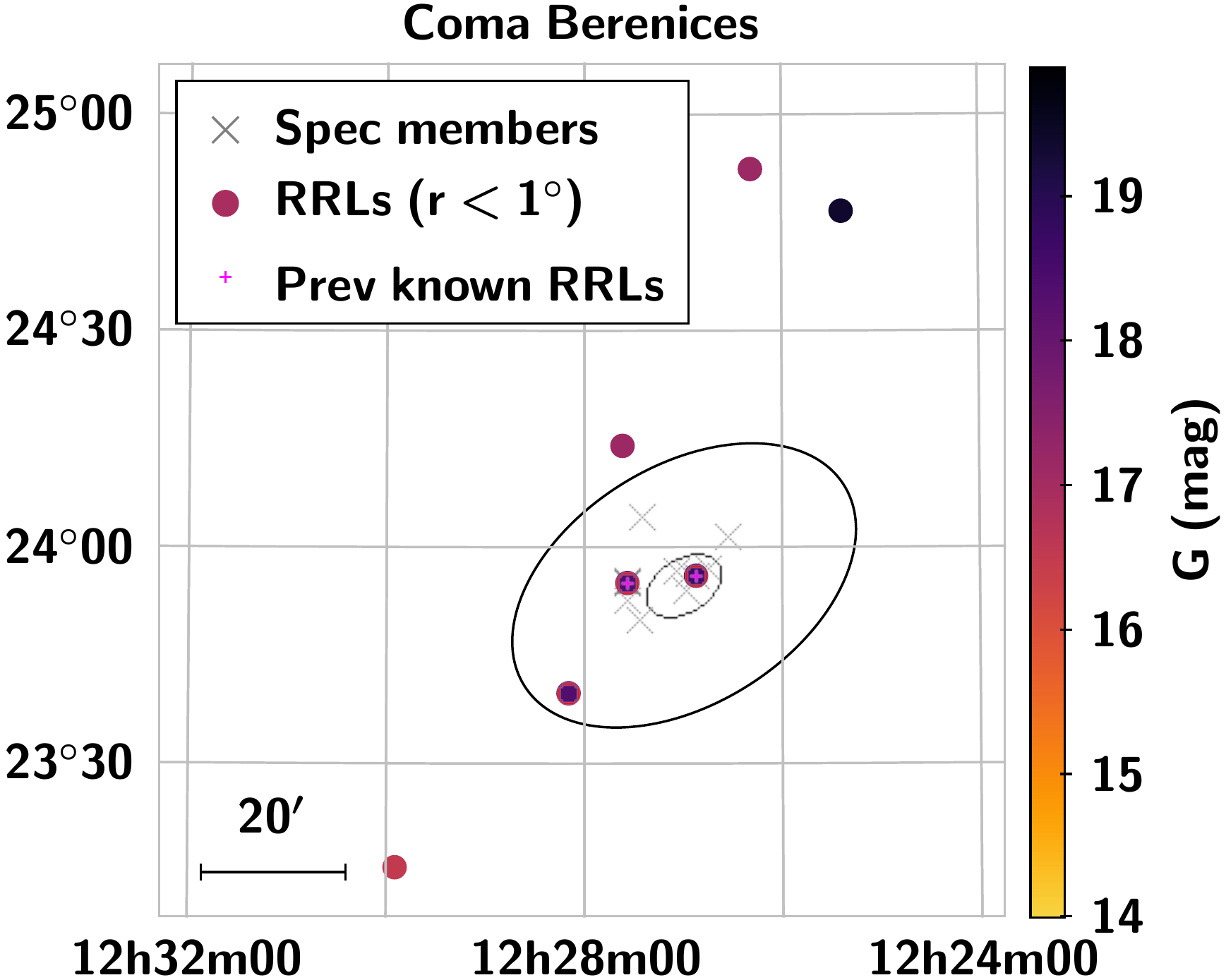}
\includegraphics[width=0.32\textwidth]{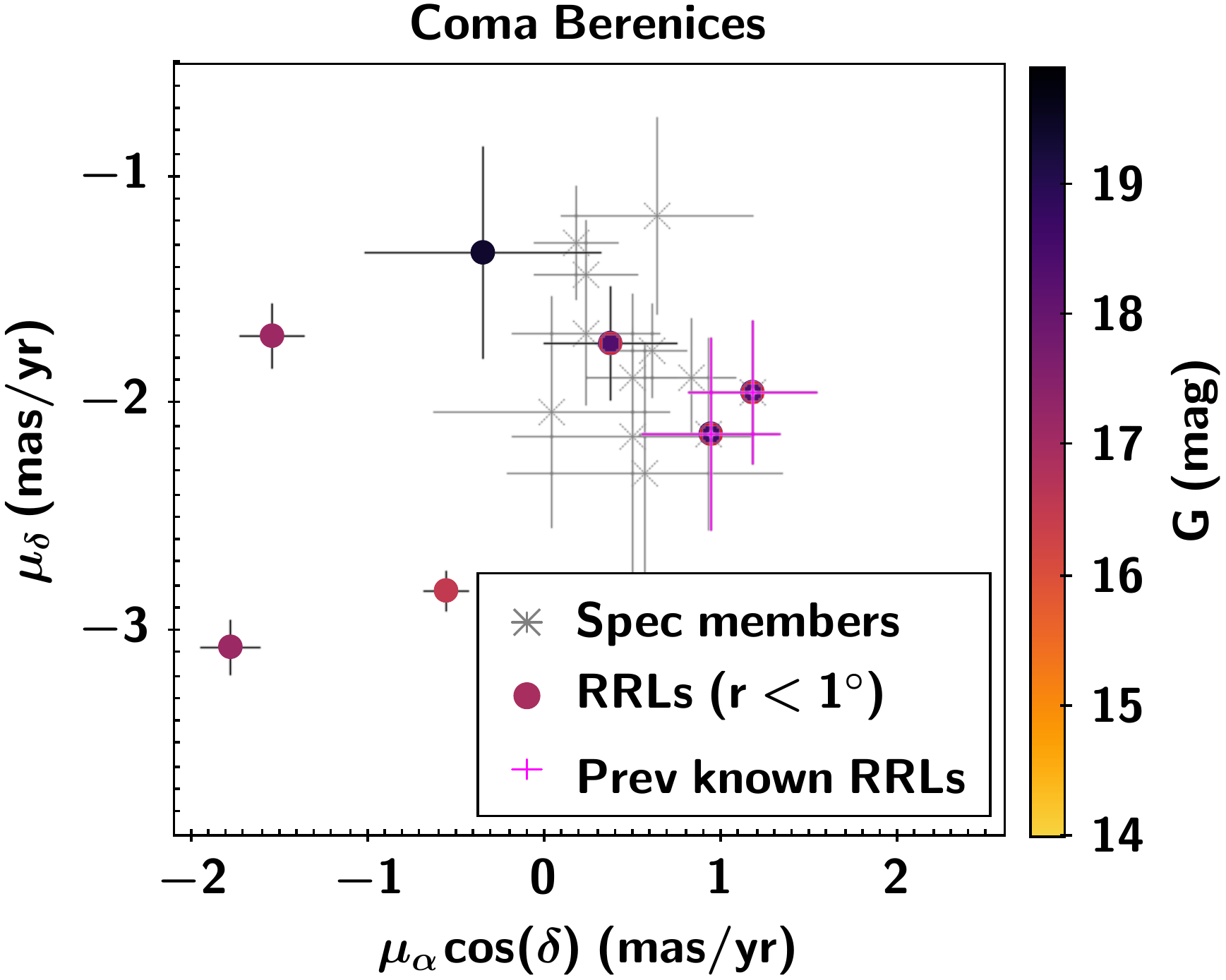}
\includegraphics[width=0.32\textwidth]{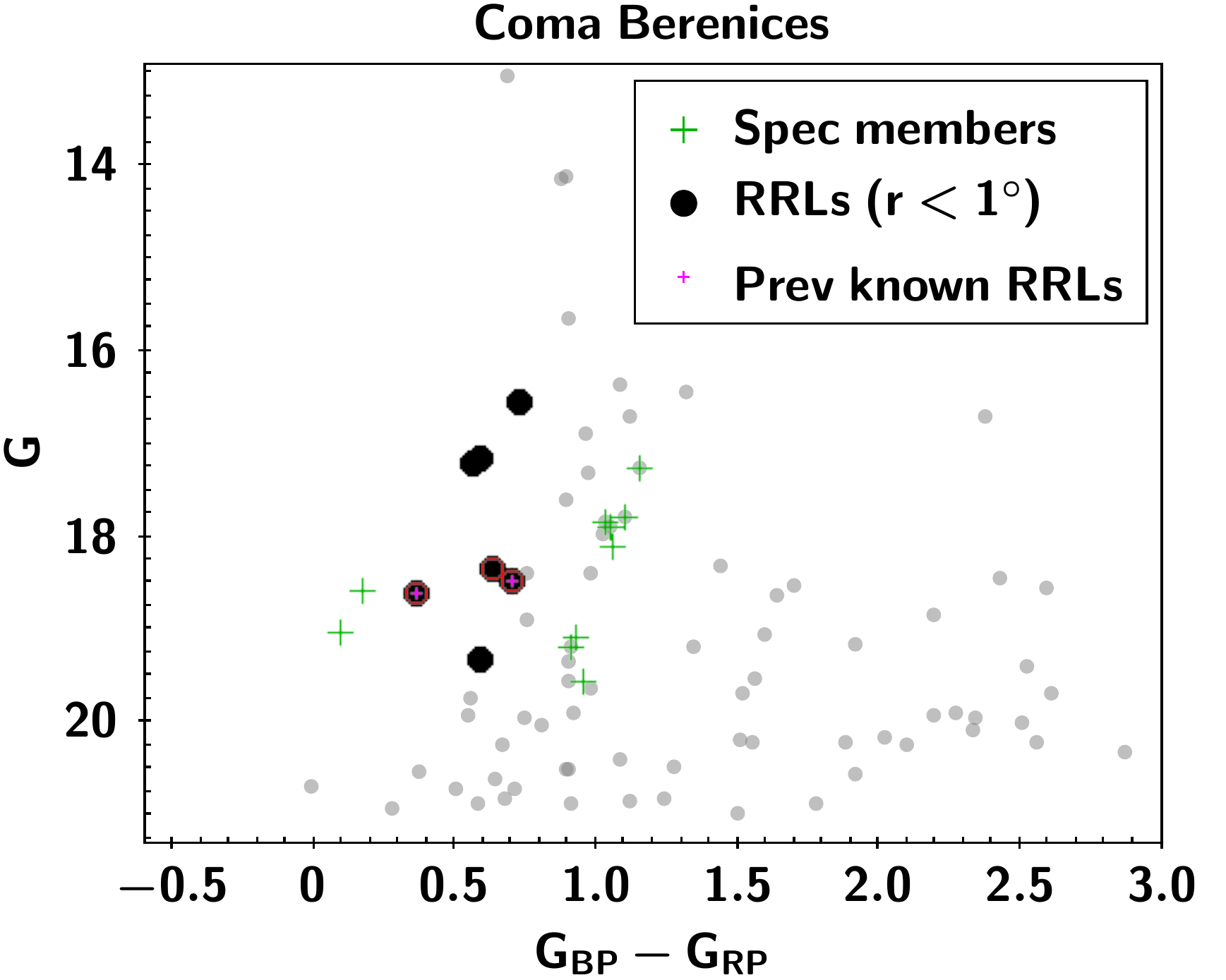}
\caption{Same as Figure~\ref{fig:UMaII} but for the ComBer UFD.}
\label{fig:CB}
\end{figure*}

\paragraph{Coma Berenices (Figure~\ref{fig:CB})}

Coma Berenices (ComBer) is also a SDSS UFD with $M_V=-4.4$. The morphology does not show signs of tidal disruption \citep{munoz18}. \citet{musella09} searched for RRLs in this galaxy and found two. We have found those two stars in Gaia, at distances of $2\arcmin$ and $8\arcmin$ from the center of ComBer. In our search we found one additional RRL in ComBer (Figure~\ref{fig:CB}), which has similar proper motion and similar magnitude as the other members. The new RRL is located at $22\arcmin$ from the center of ComBer. Although it is much farther away than the others, it is still within the tidal radius of $26\farcm 8$ measured by \citet{munoz18}. In total, ComBer then has three RRLs, two of which are {\rrab } and one is type \rrc. The periods of the two {\rrab } stars are quite similar, 0.670 d. The new star is labeled in Table~\ref{tab:RRL} as V4 since \citet{musella09} uses V3 for a short period variable detected in the field.

\begin{figure*}[htb!]
\centering
\includegraphics[width=0.32\textwidth]{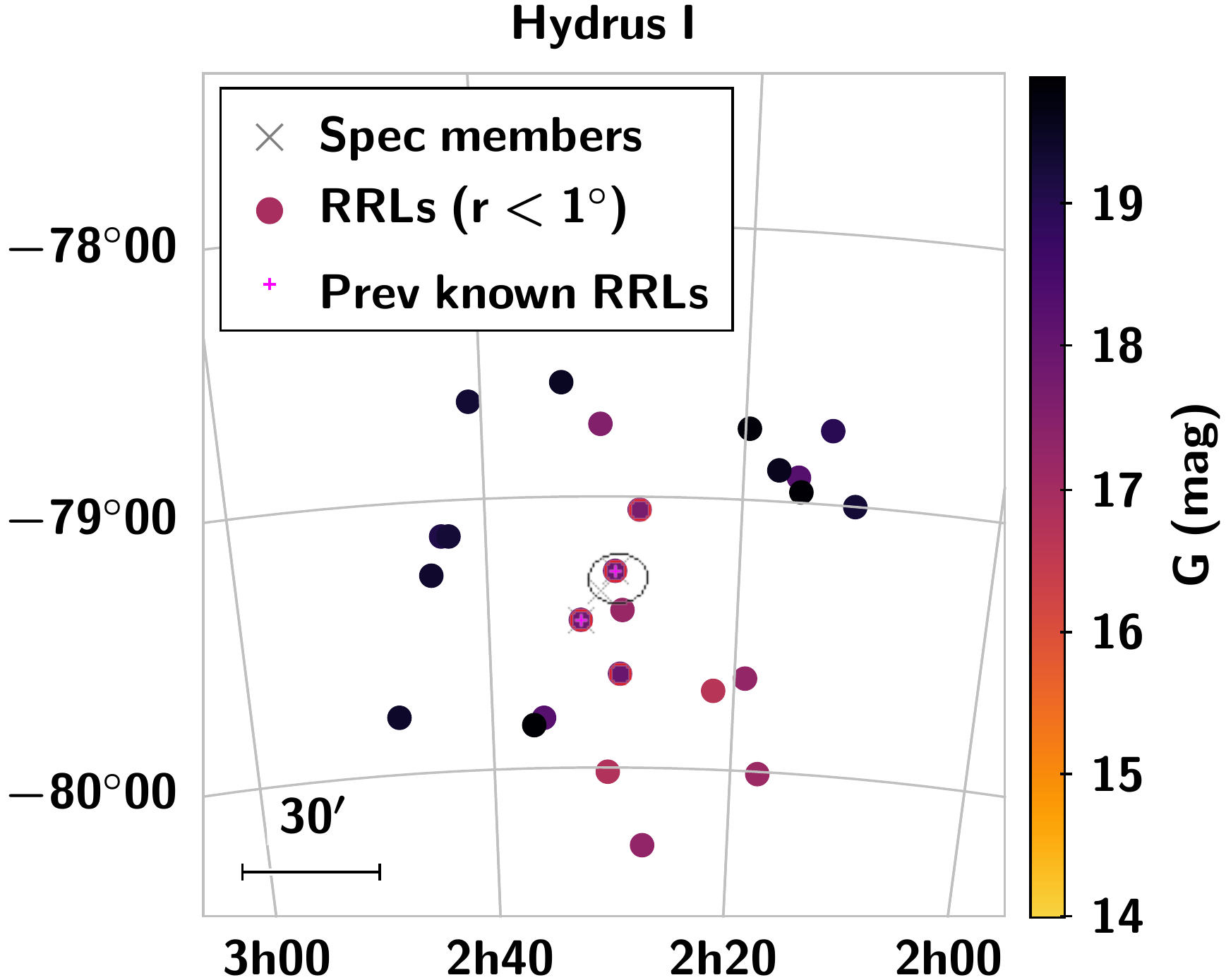}
\includegraphics[width=0.32\textwidth]{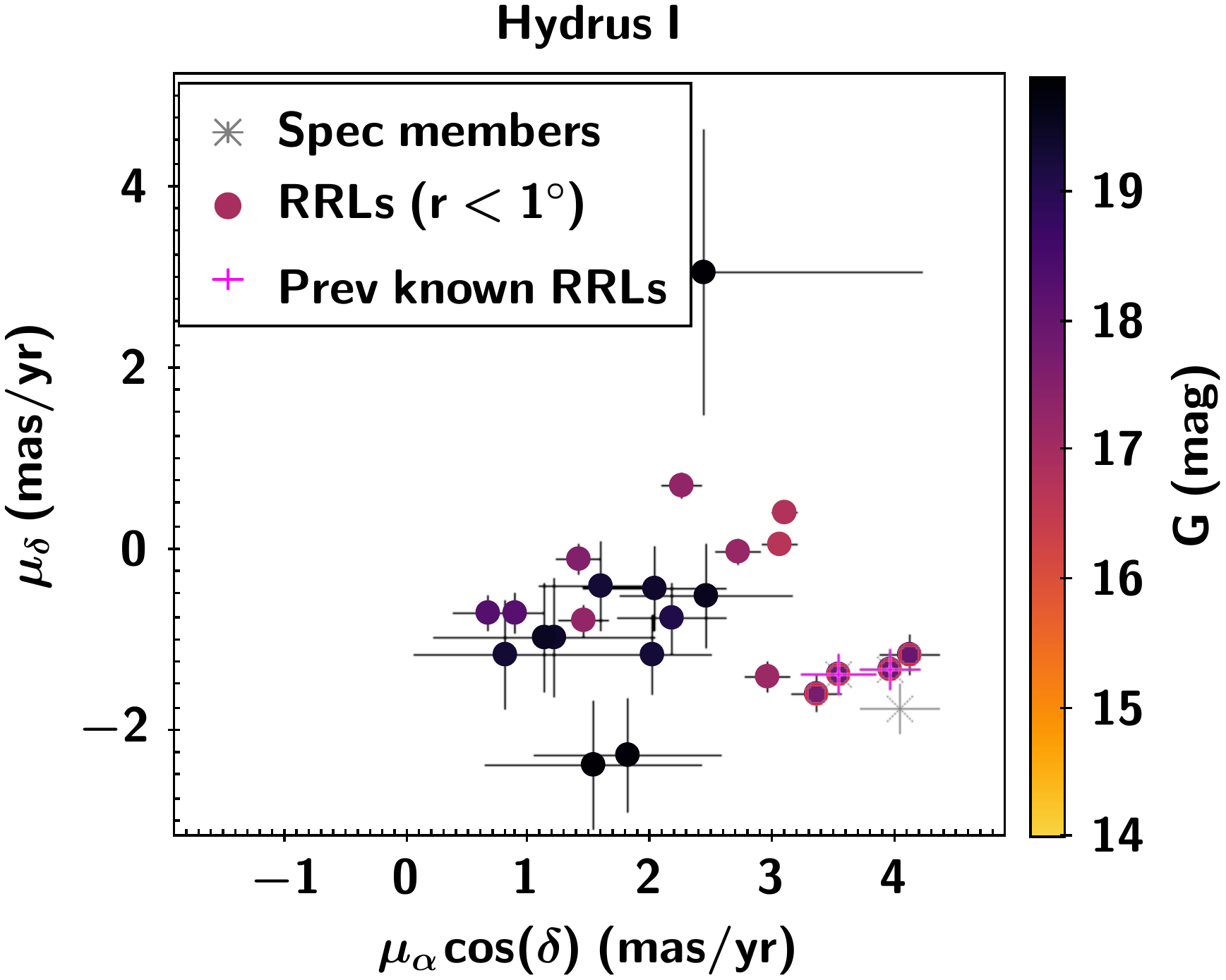}
\includegraphics[width=0.32\textwidth]{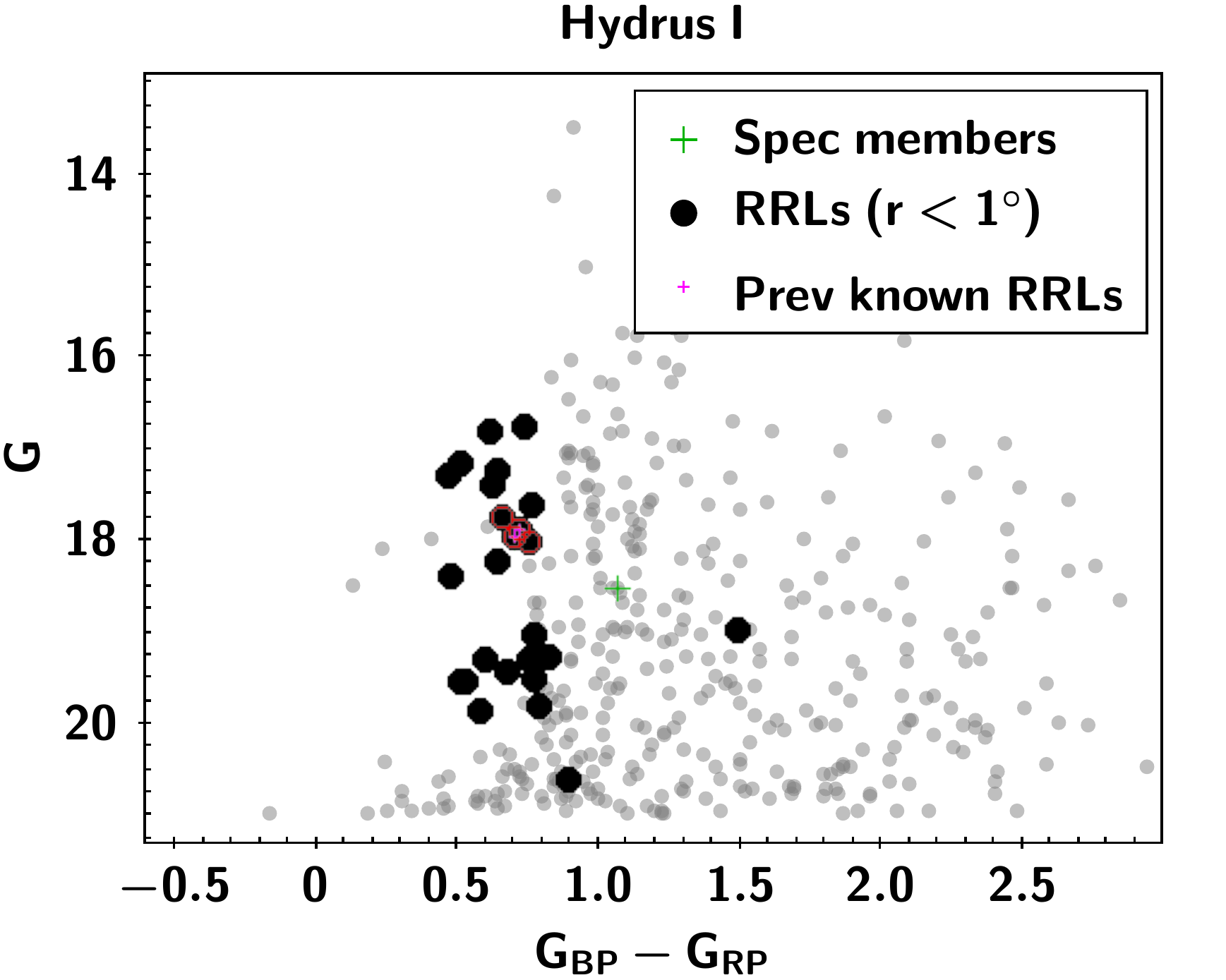}
\caption{Same as Figure~\ref{fig:UMaII} but for the Hyd I UFD. Only the half-light radius ellipse is shown.}
\label{fig:HydrusI}
\end{figure*}

\paragraph{Hydrus I (Figure~\ref{fig:HydrusI})}

Hydrus~I (Hyd~I) is a UFD satellite with $M_V\lesssim -4.7$ mag discovered by \citet{koposov18} using Dark Energy Camera (DECam) data. The kinematics for 30 stars revealed the nature of this system as a galaxy. Hyd~I is a very metal-poor galaxy with a mean metallicity [Fe/H]=$-2.5$, located between the LMC and SMC, at $\sim$28 kpc from us. Its position on the sky and its line of sight velocity, makes Hyd~I a strong candidate to be a  LMC satellite. 
The color-magnitude diagram (see Figure 2 in \citealt{koposov18}) shows a few stars in the HB. Particularly, two OGLE RRLs ($ab$ type) were associated by \citet{koposov18} to this galaxy: OGLE-SMCRRLYR-6316 and OGLE-SMC-RRLYR-6325, with periods of 0.67 and 0.73 d. In this work, we recover the two 
OGLE stars, and find two additional RRL members for Hyd~I (Figure~\ref{fig:HydrusI}). The new stars are also {\rrab } and are located in the outskirts of Hyd~I, at $16\arcmin$ and $21\arcmin$ from the center, which means that the RRLs extend up to $3.2\, r_h$. We renamed here the two stars reported by \citet{koposov18} as V1 and V2, and assigned V3 and V4 to the new RRLs.

\begin{figure*}[htb!]
\centering
\includegraphics[width=0.32\textwidth]{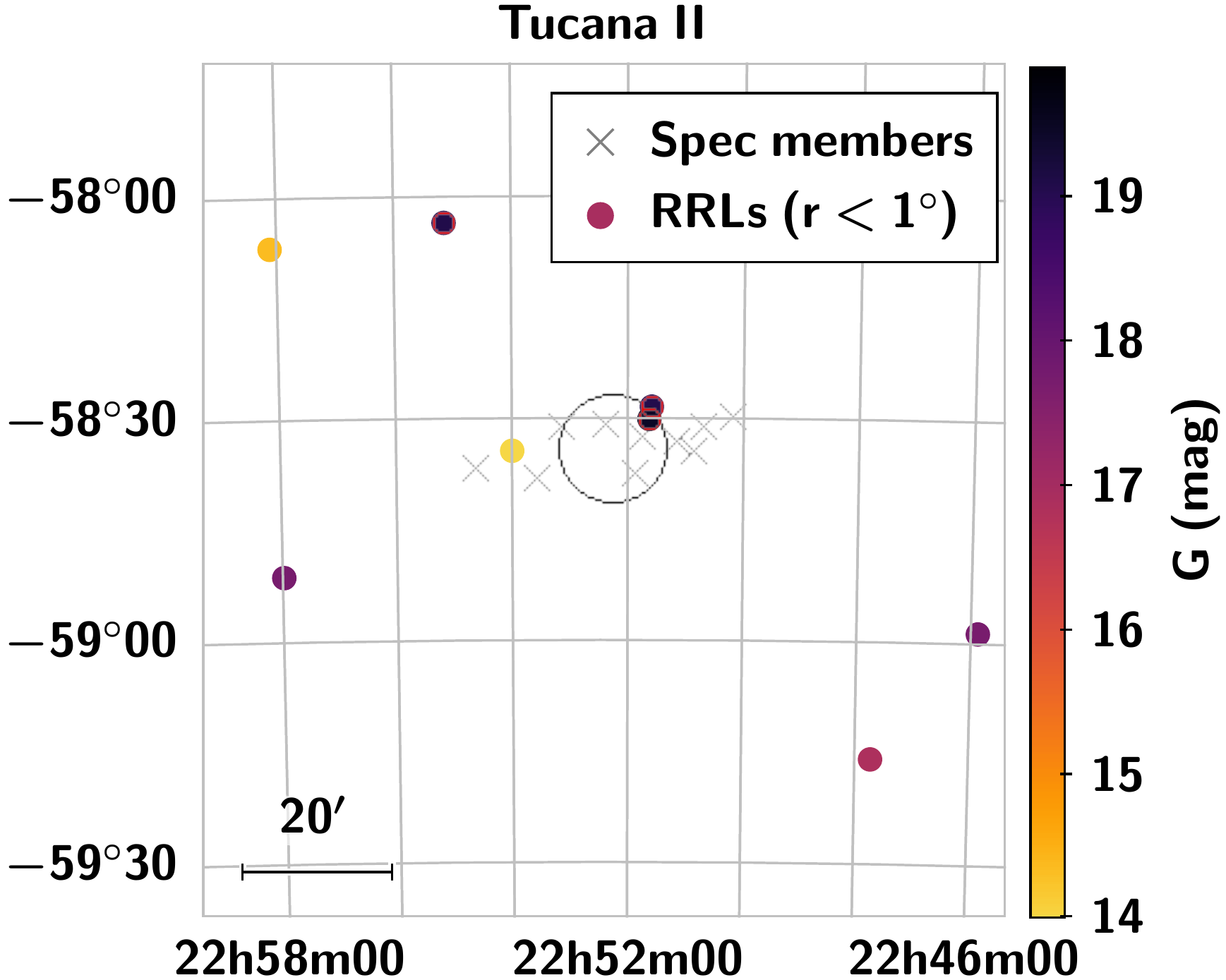}
\includegraphics[width=0.32\textwidth]{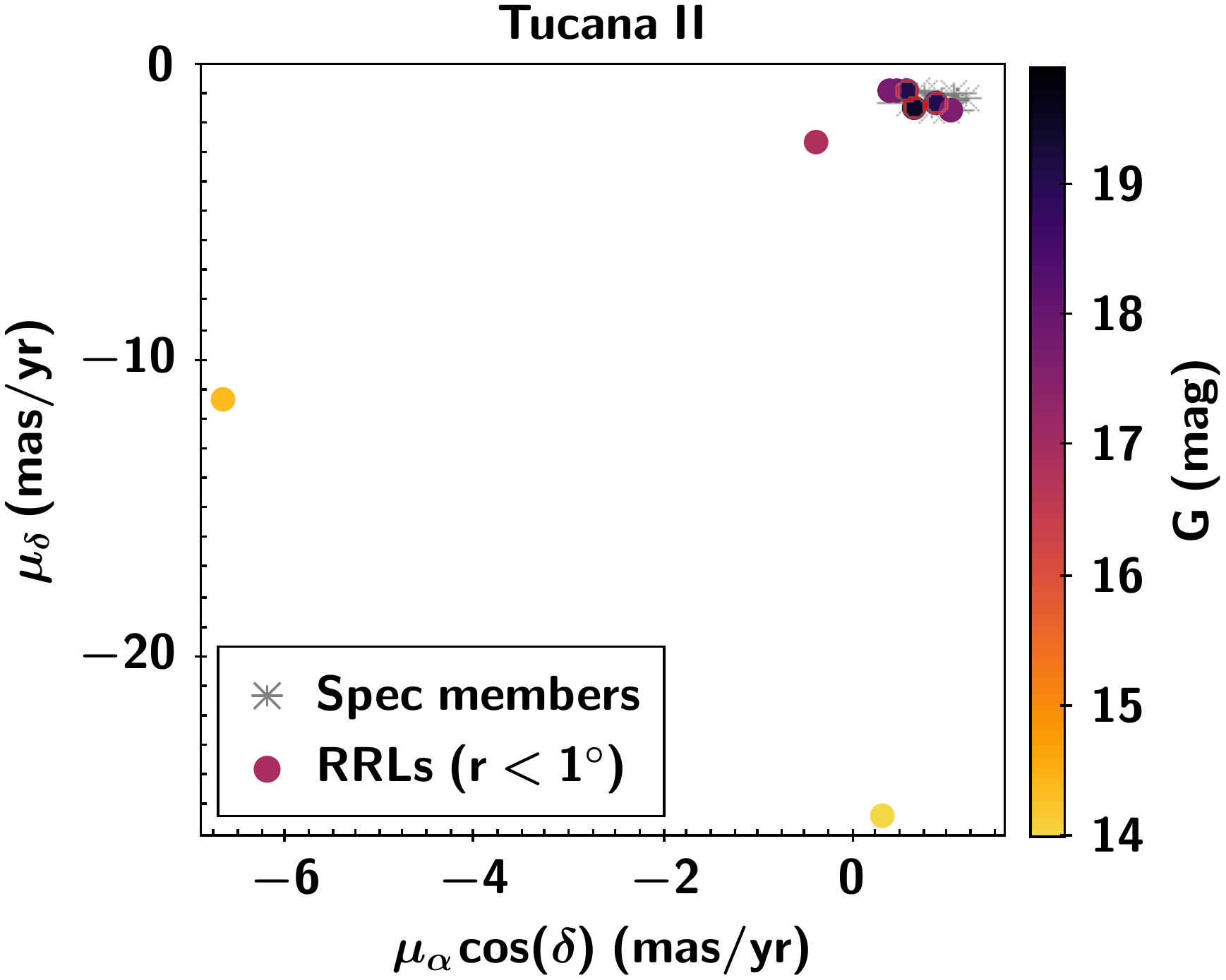}
\includegraphics[width=0.32\textwidth]{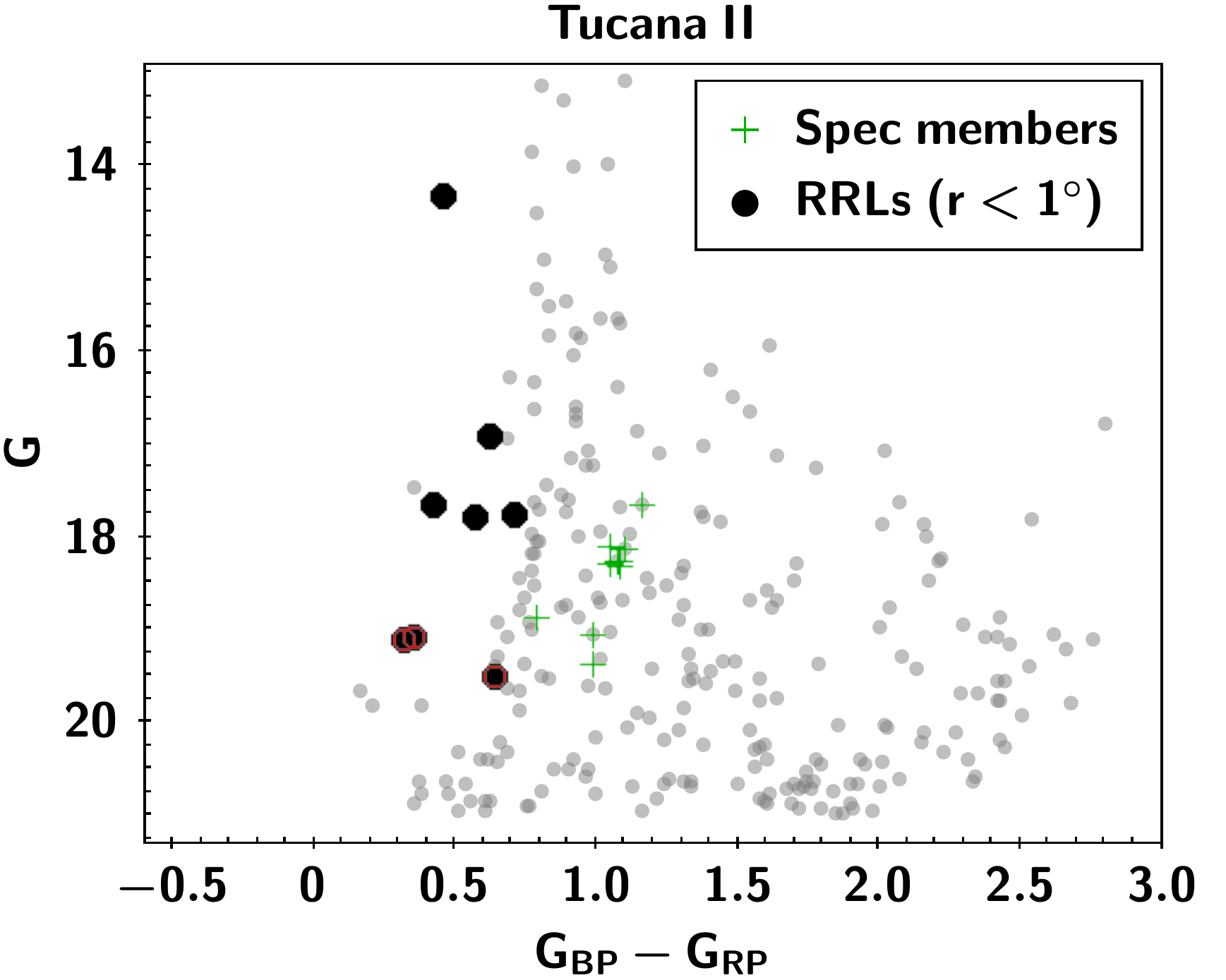}
\caption{Same as Figure~\ref{fig:UMaII} but for the Tuc II UFD. Only the half-light radius ellipse is shown.}
\label{fig:TucII}
\end{figure*}

\paragraph{Tucana II (Figure~\ref{fig:TucII})}

Tucana~II (Tuc~II) is a UFD ($M_V = -3.8$ mag) detected in DES by two independent groups \citep{koposov15,bechtol15}. Its large physical size classifies the system as a dwarf galaxy. It is located close to the LMC (at $\sim$ 19 kpc) and at $\sim$32 kpc from us. Due to its proximity to the LMC, it is considered a likely LMC satellite.
The color-magnitude diagram of Tuc~II (see e.g., Figure 6 in \citealt{bechtol15}) reveals some stars located at the HB. We report here for the first time the detection of three RRLs as members of Tuc~II, one is a {\rrab } and the other two are \rrc. The three stars have mean $G$ magnitudes between 19.0 and 19.5, and are located at $5\arcmin$, $6\arcmin$, and $39\arcmin$. The last one is at $5\; r_h$. It is not clear if that would make this an extra-tidal star. 

\subsubsection{UFD galaxies with new RRLs members including extra-tidal candidates}

The wide search we made in the Gaia catalog allowed us to identify possible extra-tidal RRLs. These stars share the same proper motion and distance as the galaxies, but they are located far from their centers. Confirmation via radial velocities is desirable. In this category we have three galaxies, Boo I, Boo III and Sgr II, which seem to have extra-tidal stars in addition to members within their tidal radius. We also found three cases, Tuc III, Eridanus III (Eri III), Reticulum III (Ret III), in which the galaxy itself does not contain any RRLs, but for which we can associate extra-tidal candidates. However, we caution that both Eri III and Ret III are distant galaxies and consequently their RRLs in Gaia are faint and the lightcurves noisy. 

Extra-tidal candidates are marked with an asterisk in Table~\ref{tab:RRL}.

\begin{figure*}[htb!]
\centering
\includegraphics[width=0.32\textwidth]{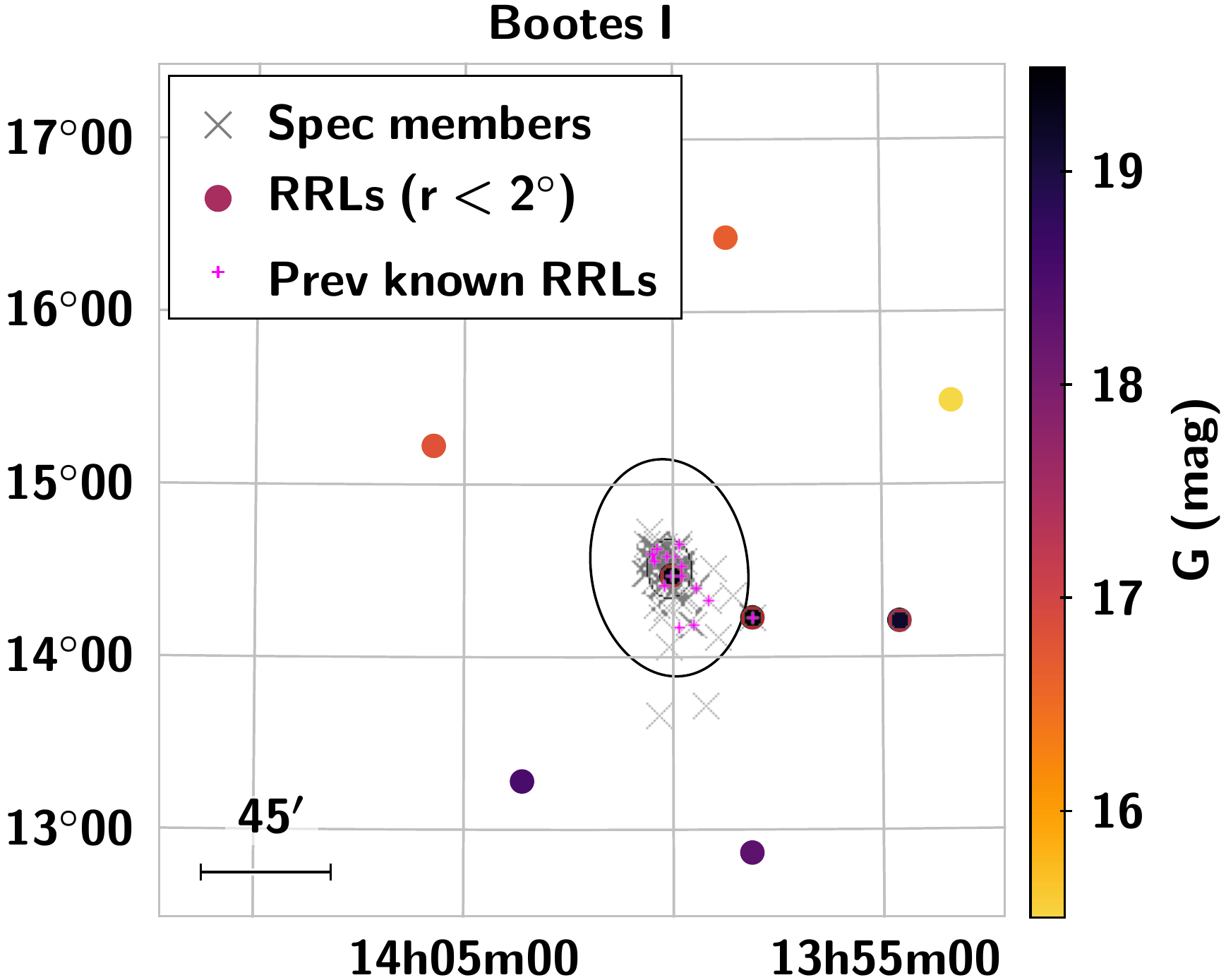}
\includegraphics[width=0.32\textwidth]{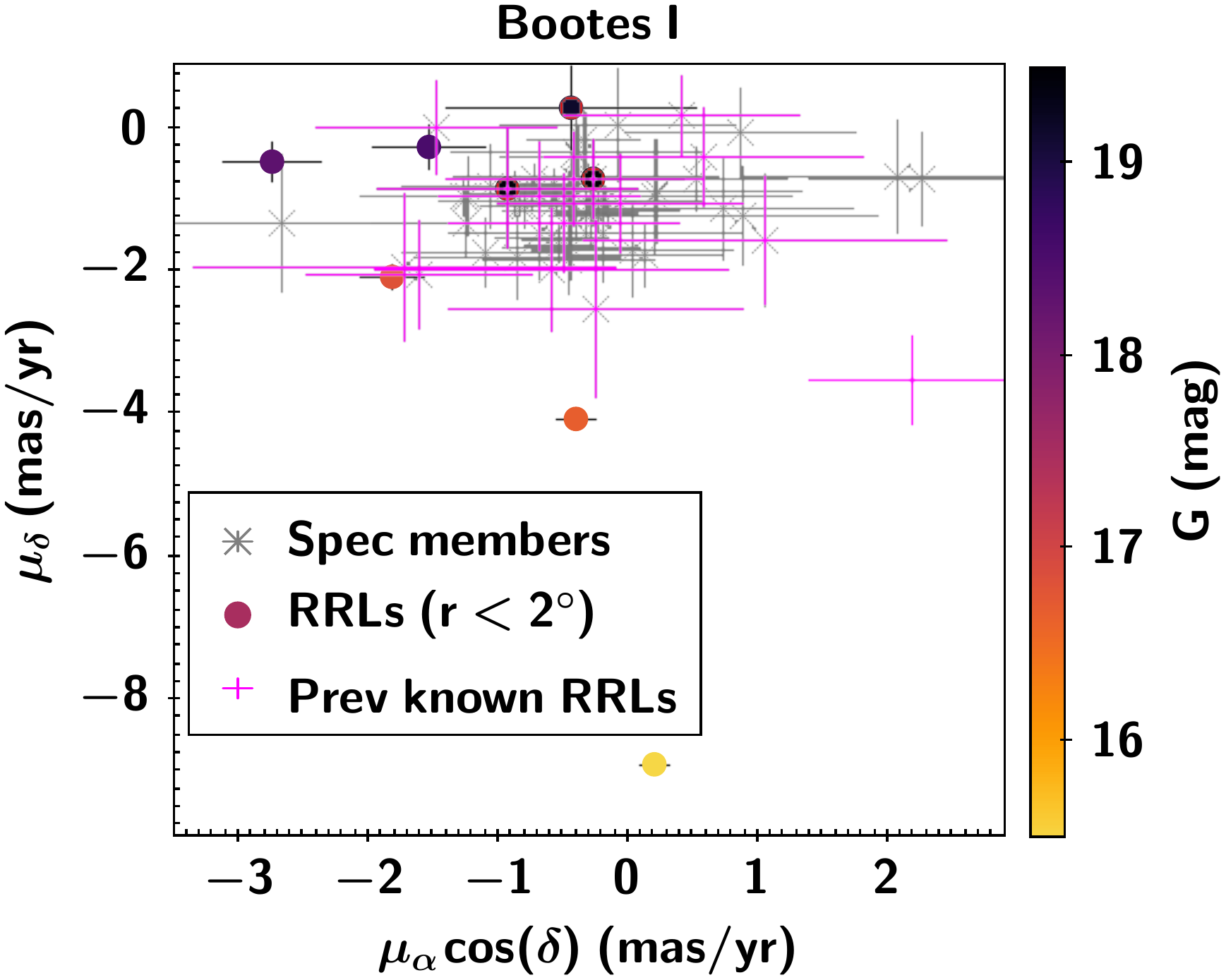}
\includegraphics[width=0.32\textwidth]{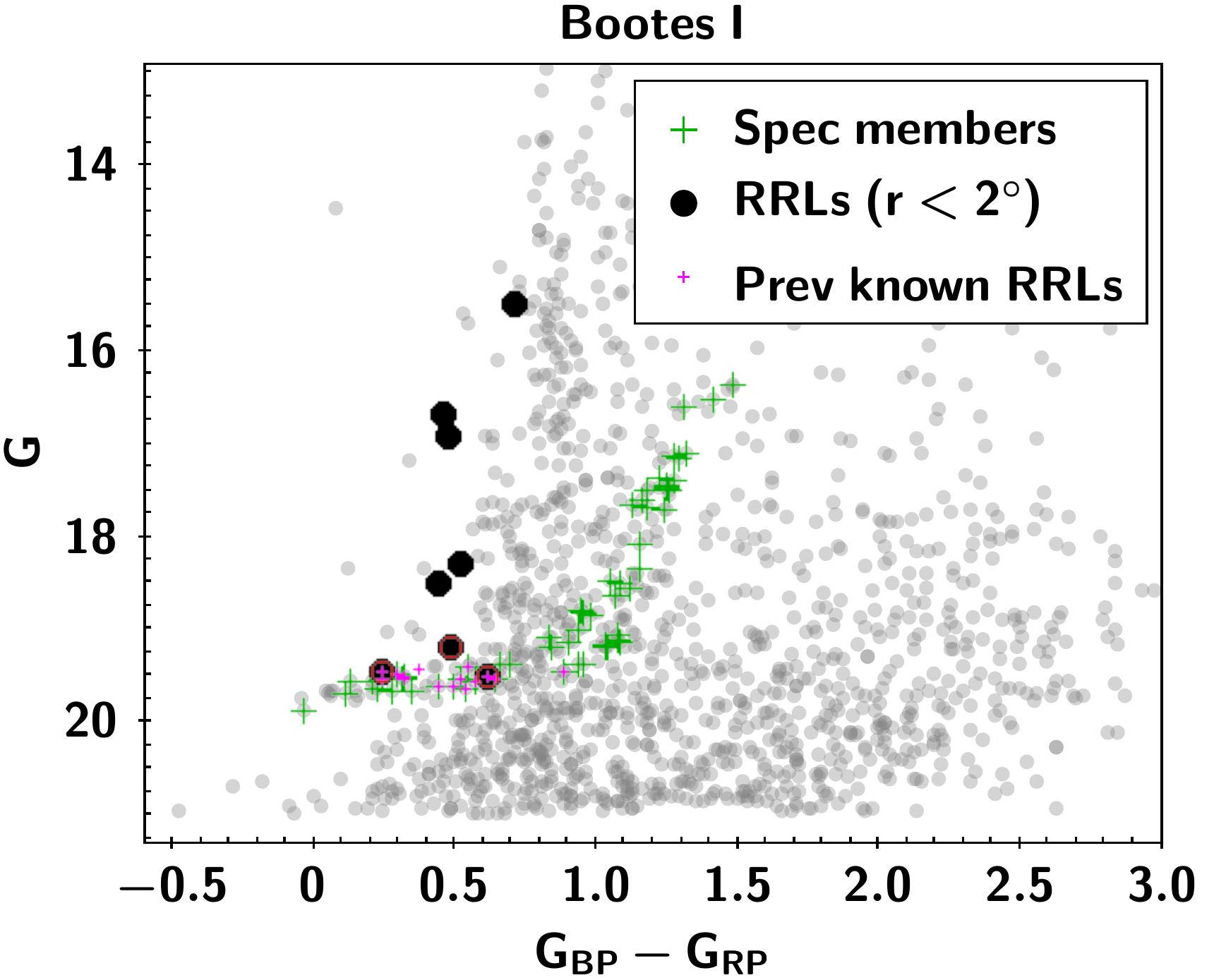}
\caption{Same as Figure~\ref{fig:UMaII} but for the Boo I UFD.}
\label{fig:BooI}
\end{figure*}

\begin{figure*}[htb!]
\centering
\includegraphics[width=0.32\textwidth]{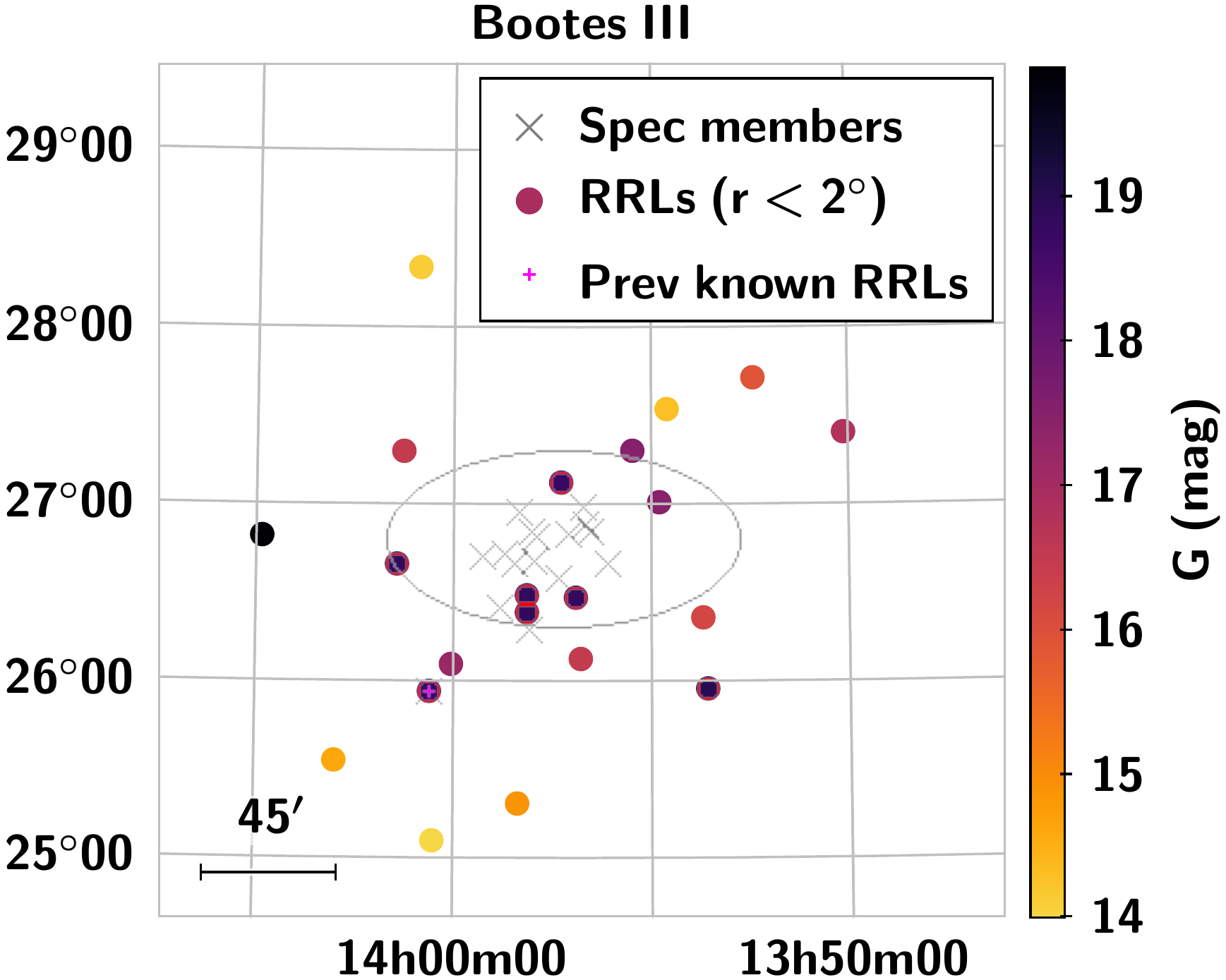}
\includegraphics[width=0.32\textwidth]{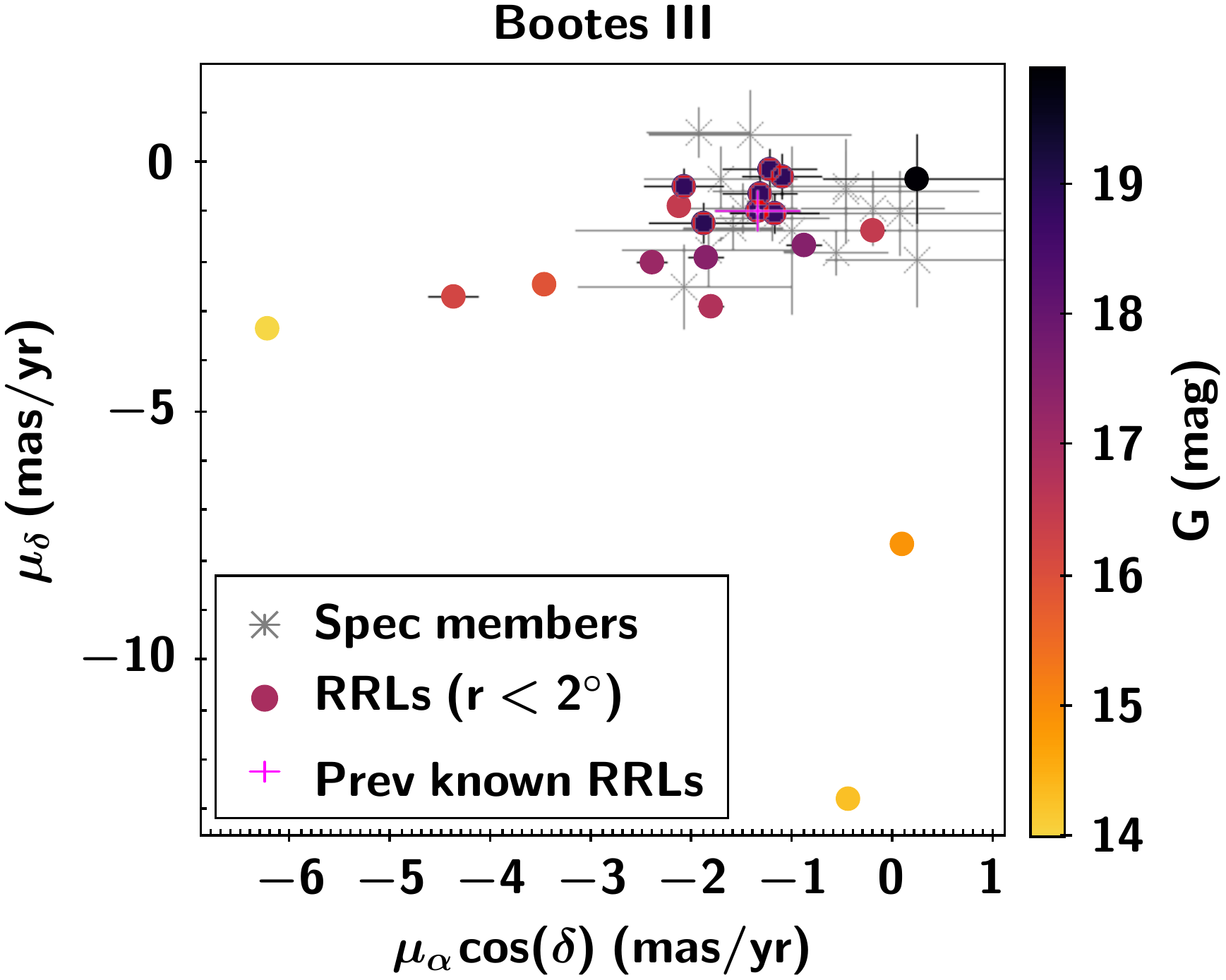}
\includegraphics[width=0.32\textwidth]{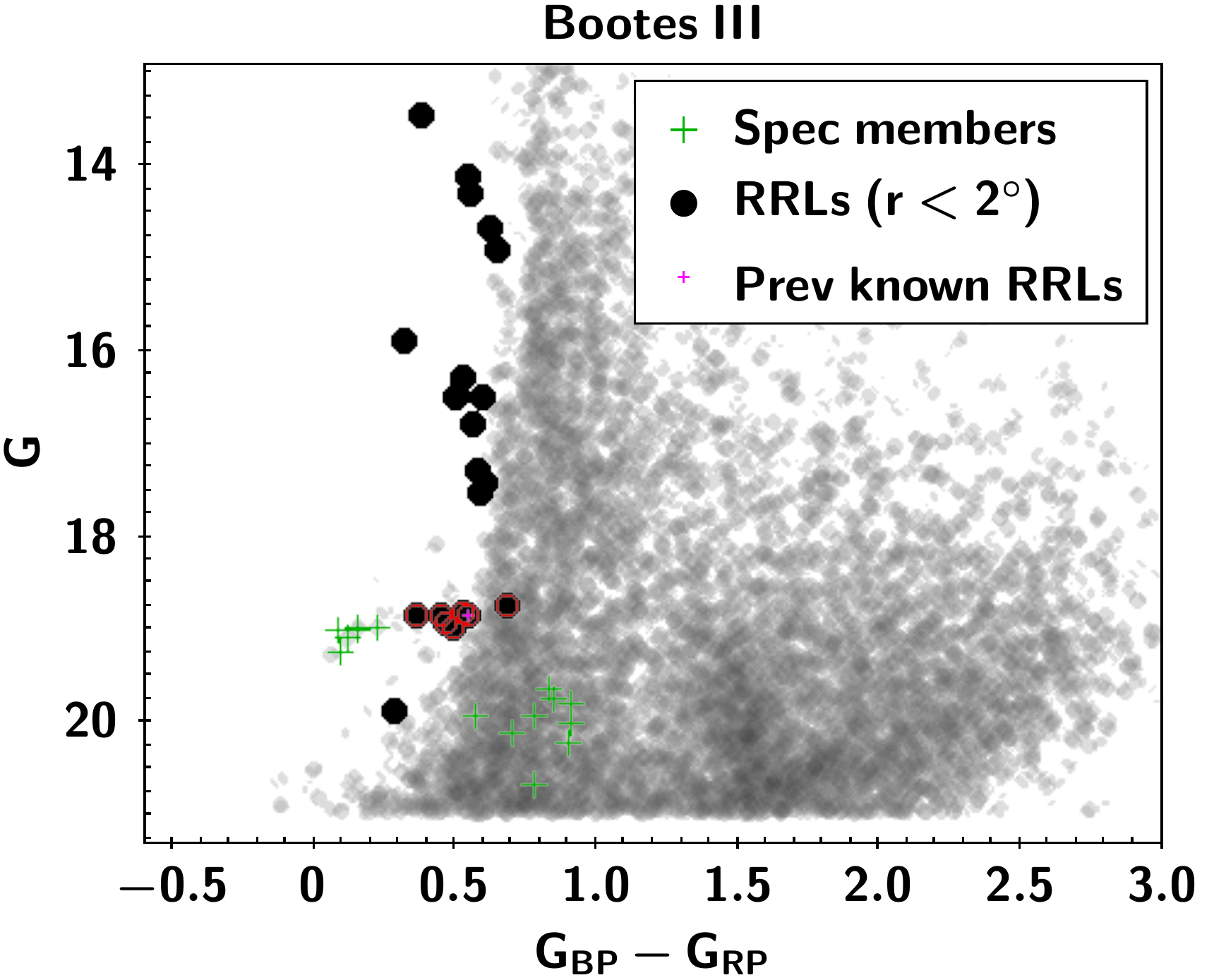}
\caption{Same as Figure~\ref{fig:UMaII} but for the Boo III UFD. Only the half-light radius ellipse is shown.}
\label{fig:BooIII}
\end{figure*}

\begin{figure*}[htb!]
\centering
\includegraphics[width=0.32\textwidth]{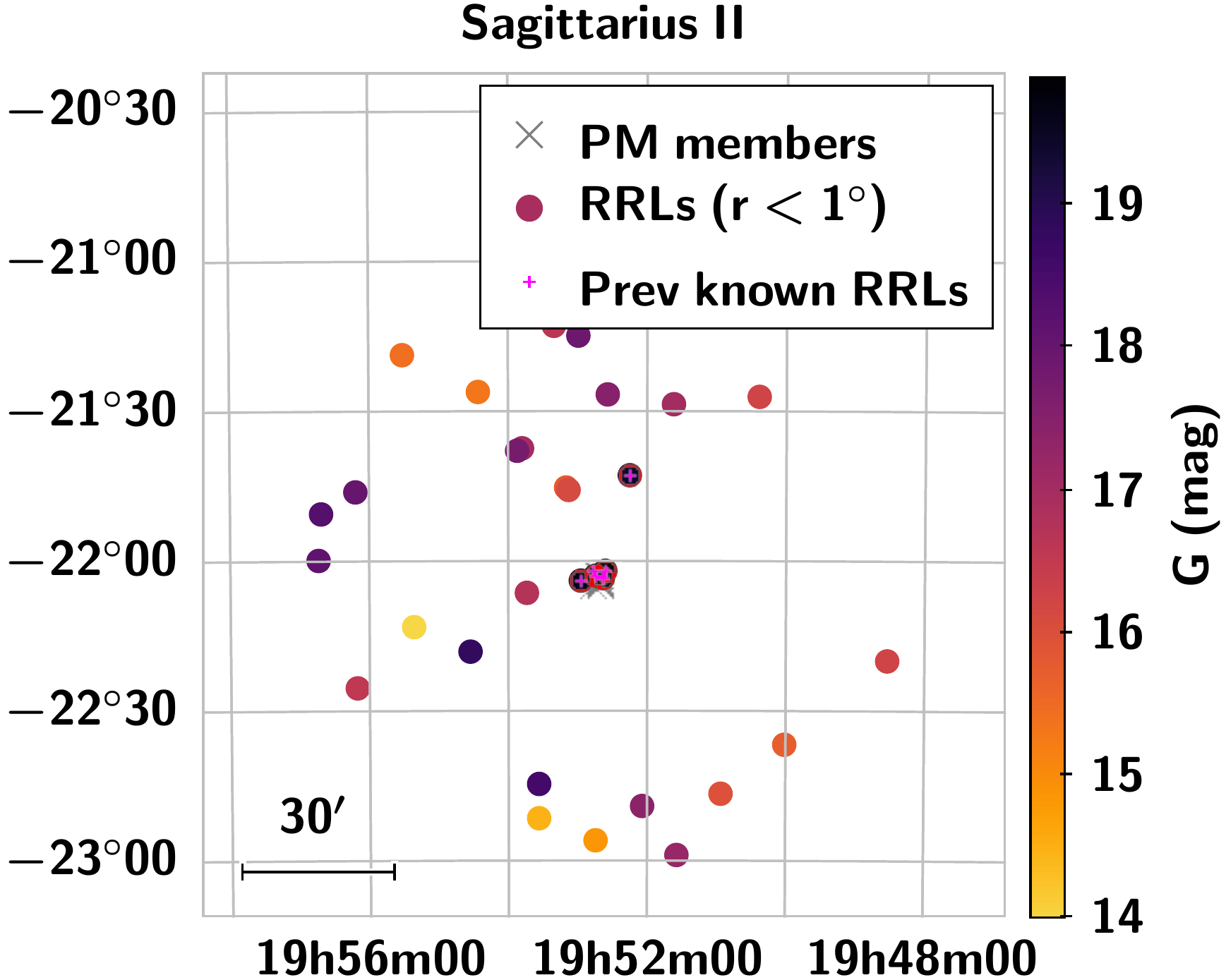}
\includegraphics[width=0.32\textwidth]{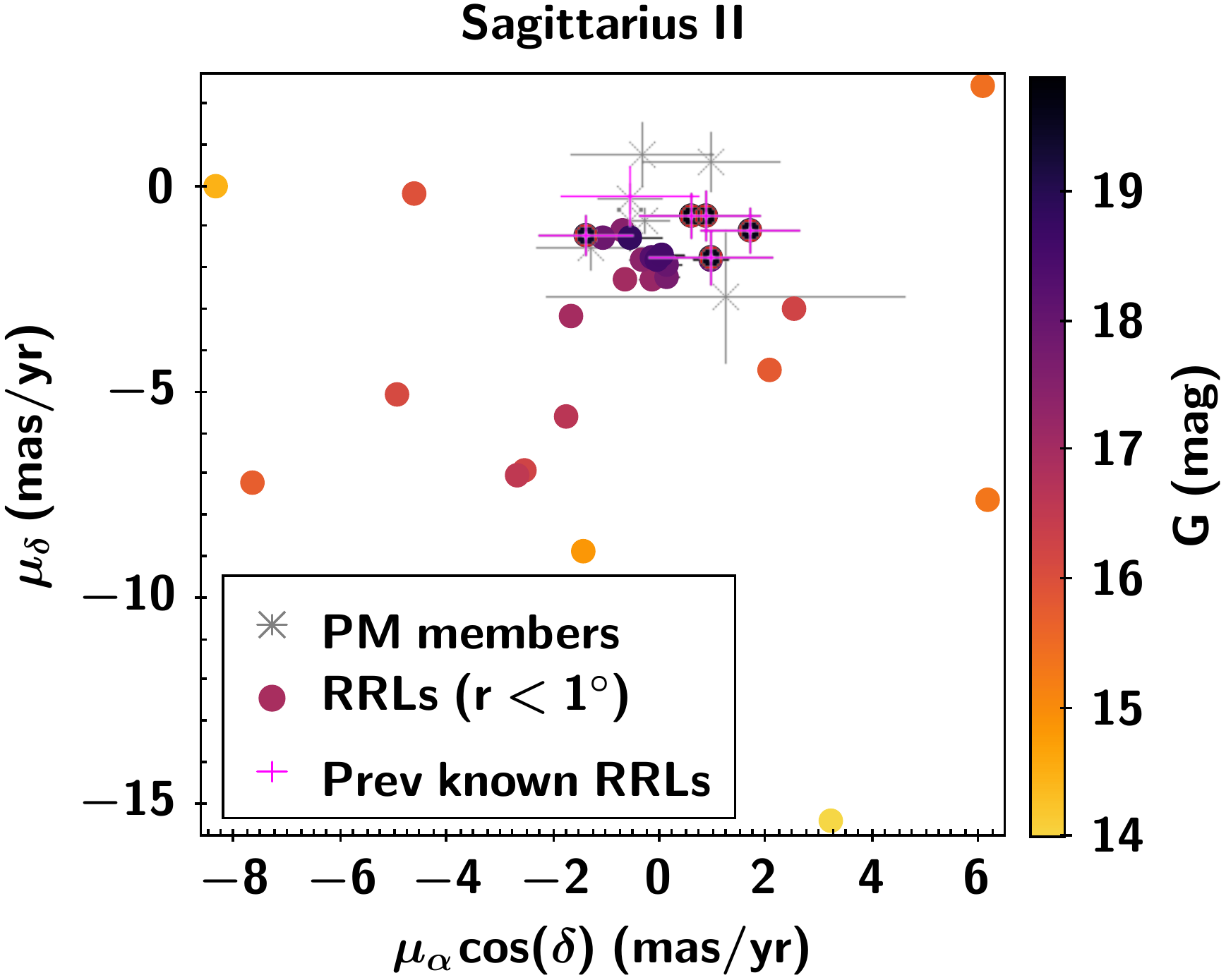}
\includegraphics[width=0.32\textwidth]{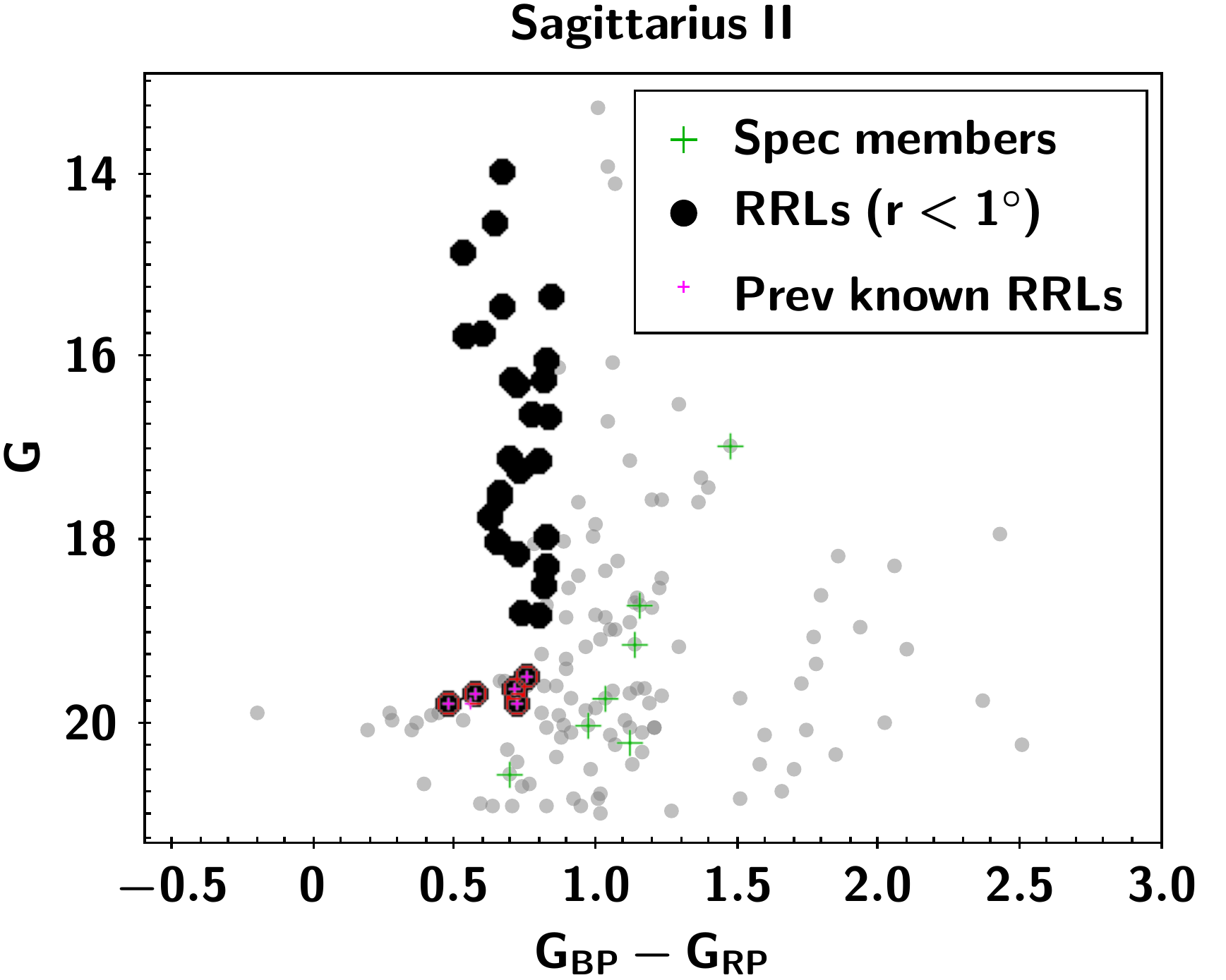}
\caption{Same as Figure~\ref{fig:UMaII} but for the Sgr II UFD. Only the half-light radius ellipse is shown.}
\label{fig:SgrII}
\end{figure*}

\paragraph{Bootes I (Figure~\ref{fig:BooI})}

Boo I is the brightest of the sample of UFDs closer to 100 kpc with $M_V=-6.02$. It has a rich population of RRLs. \citet{siegel06} detected 15 RRLs in this galaxy \citep[see also][]{dallora06}. \citet{siegel06} assigned a Oo~II classification to this galaxy. The completeness of the Gaia DR2 catalog must be particularly low in this part of the sky since it only recovers two out of the 15 stars previously known. In this galaxy we used a search radius of $2\degr$ since this is a large galaxy, with a tidal radius of $38\arcmin$ \citep{munoz18}. With this extended area, we were able to detect one additional star which has magnitude and proper motions in agreement with other members (Figure~\ref{fig:BooI}). This star (named V16 in Table~\ref{tab:RRL}), as well as the previously known V11, seem to be extra-tidal stars. The new star is a type c star, which brings the final census for this galaxy to seven {\rrab } and nine \rrc. Using proper motion information from the general Gaia DR2 catalog for the unrecovered RRLs in this galaxy, we confirm that all but one star are proper motion members of Boo I. Star V7 in \citet{siegel06}, a type c star, seems discrepant, although error bars are large at the magnitude of the HB of this galaxy.

\paragraph{Bootes III (Figure~\ref{fig:BooIII})}

Boo III is a UFD ($M_V = -5.8$ mag) discovered in SDSS by \citep{grillmair09}. Boo~III is a disrupted dwarf galaxy and could be the progenitor of the Styx stream \citep{grillmair09}. Its large velocity dispersion together with its morphological parameters and orbit suggest that
Boo~III is a UFD nearing complete destruction \citep{carlin09}. Its half-light radius is large, $1\degr$. The search in this galaxy was done within a $2\degr$ radius from the center of Boo III.

\citet{sesar14} detected one RRL belonging to Boo~III. With Gaia DR2, we recognized seven RRLs (Figure~\ref{fig:BooIII}), including the one previously known, within our search area around Boo III, with coherent proper motions and similar magnitudes. Boo III has then four {\rrab }, and three \rrc. The mean periods of the {\rrab } is 0.65 d, which agrees with being a Oo~II system. In Table~\ref{tab:RRL} we renamed the star discovered by \citet{sesar14} as V1, and assigned V2 to V7 to the new detections. Two of the new RRLs are located beyond the tidal radius of Boo III, which is not surprising since it is known this galaxy is disrupting.

\paragraph{Sagittarius II (Figure~\ref{fig:SgrII})}

Sgr II is the third brightest galaxy in our sample with $M_V=-5.7$. It is believed that Sgr II is a case of a satellite of a satellite since it has an orbit similar to the classical Sgr dSph \citep{longeard20}, which we know is disrupting. Sgr II was searched for RRLs by \citet{joo19}. They found six RRLs in the field, but associated only five of them to Sgr II because one of them, V4, was too far away from the center of the galaxy. Gaia recovered five out of the six stars in \citeauthor{joo19}, including V4. As seen in Figure~\ref{fig:SgrII}, because of the similarity in magnitude and the agreement between the proper motions of V4 and the rest of the members of Sgr II, we believe V4 should be indeed an extra-tidal RRLs of Sgr II.

\begin{figure*}[htb!]
\centering
\includegraphics[width=0.32\textwidth]{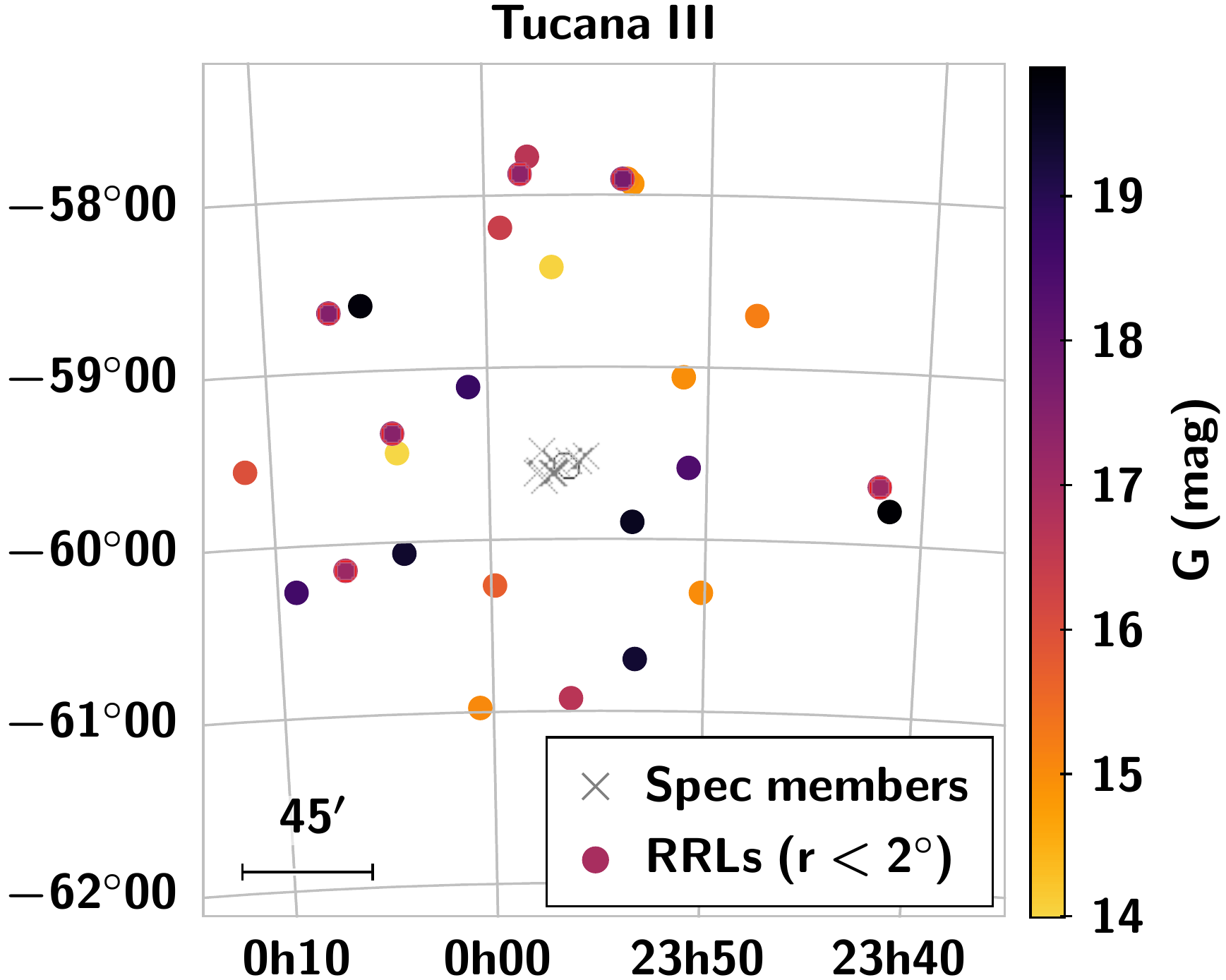}
\includegraphics[width=0.32\textwidth]{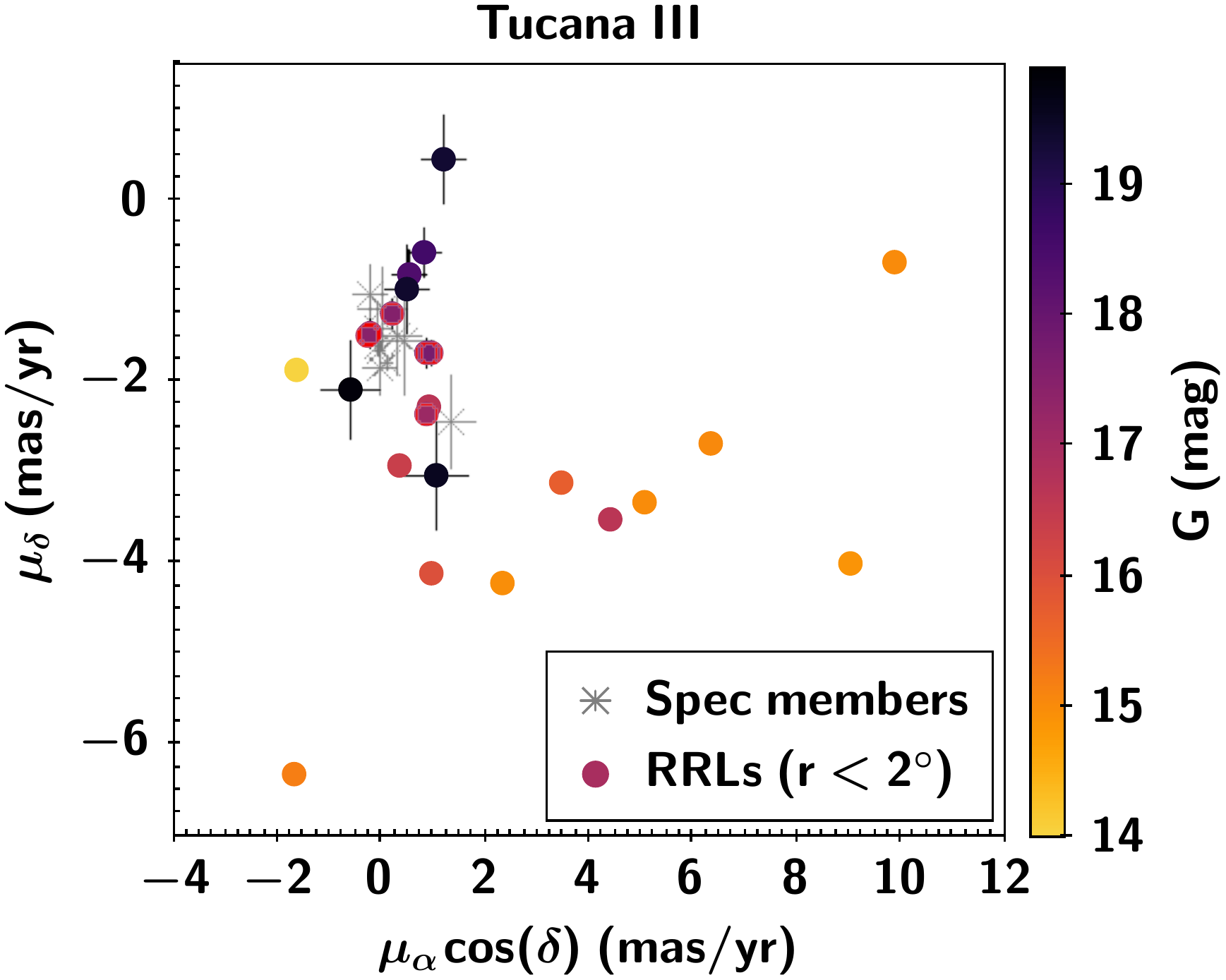}
\includegraphics[width=0.32\textwidth]{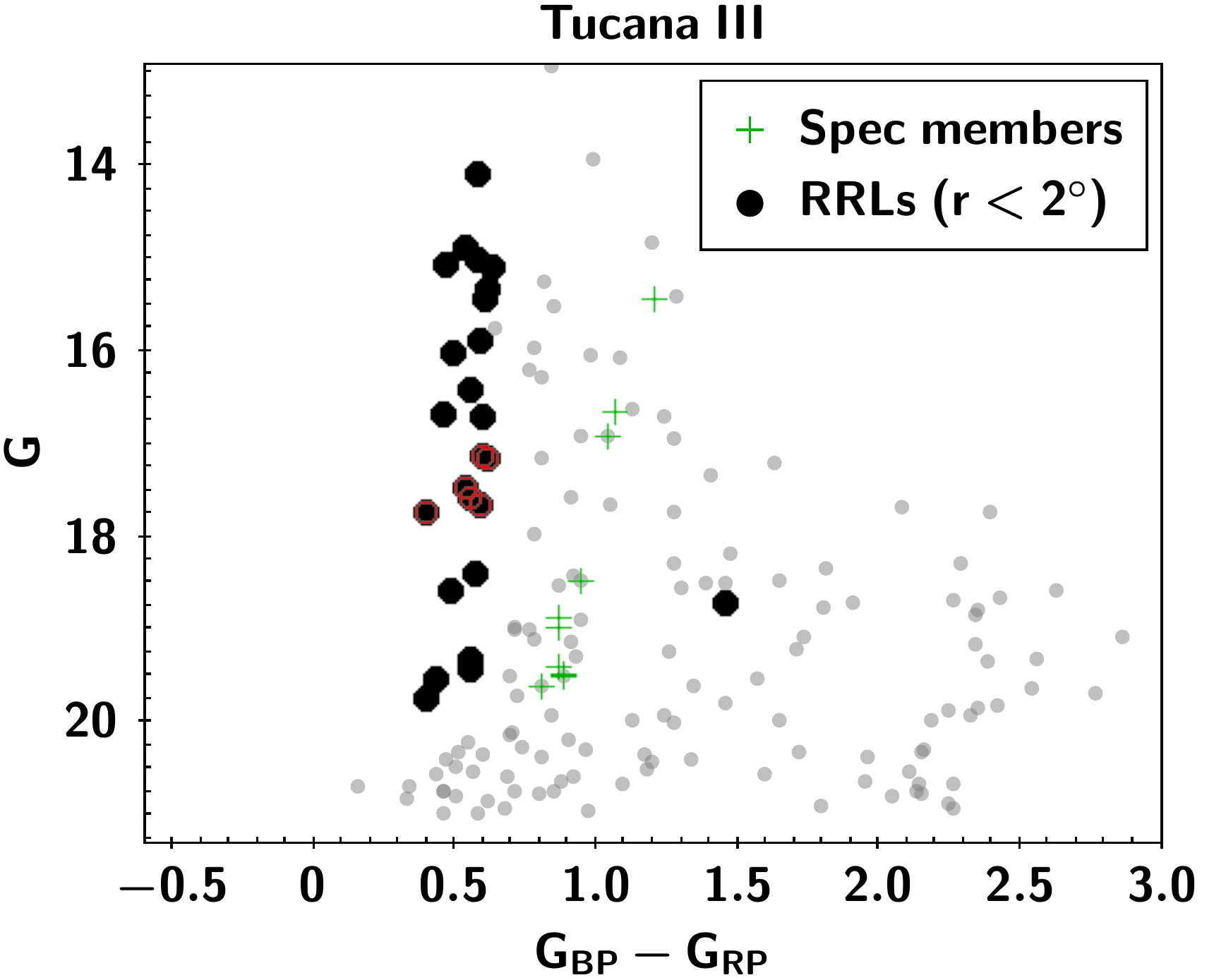}
\caption{Same as Figure~\ref{fig:UMaII} but for the Tuc III UFD. Only the half-light radius ellipse is shown.}
\label{fig:TucIII}
\end{figure*}

\paragraph{Tucana III (Figure~\ref{fig:TucIII})}

The most interesting case of the UFDs with extra-tidal candidates is that of Tuc III. Tuc III is known to be disrupting since it shows clear tidal tails that extend to a few degrees from the galaxy \citep{drlica15,shipp18,li18}. Tuc III is a relatively close system, at only 23 kpc from the Sun. Our search for Gaia RRLs in Tuc III resulted in none in the galaxy itself but six at distances between $60\arcmin$ and $110\arcmin$ from the center of Tuc III, and with a narrow range of magnitudes between 17.1-17.7 (Figure~\ref{fig:TucIII}). Surprisingly, those star do not seem to follow the tidal tails but are distributed all around the galaxy. If radial velocities confirm the association of those RRLs to Tuc III, all the RRLs population in Tuc III would have been stripped off the galaxy.

\begin{figure*}[htb!]
\centering
\includegraphics[width=0.32\textwidth]{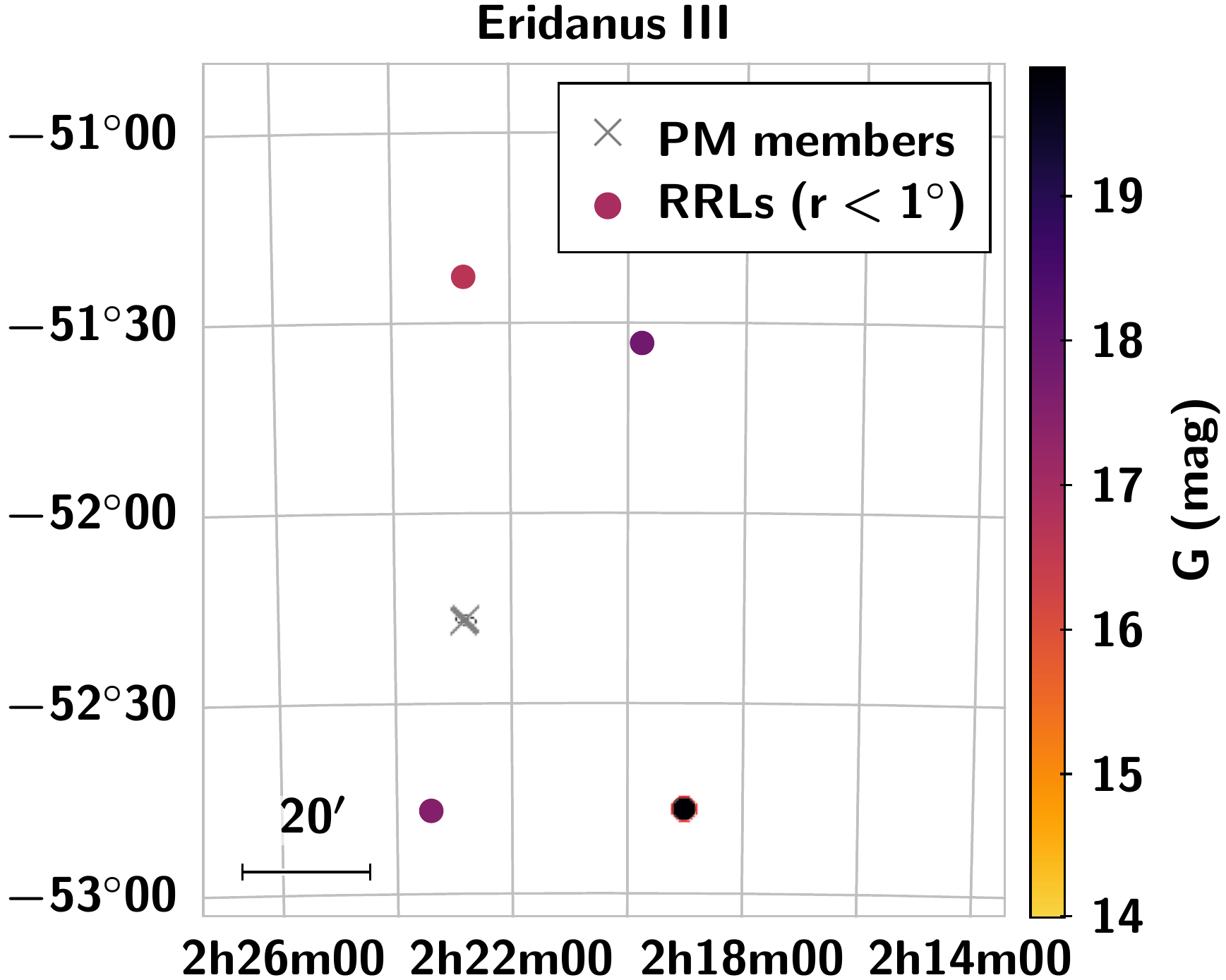}
\includegraphics[width=0.32\textwidth]{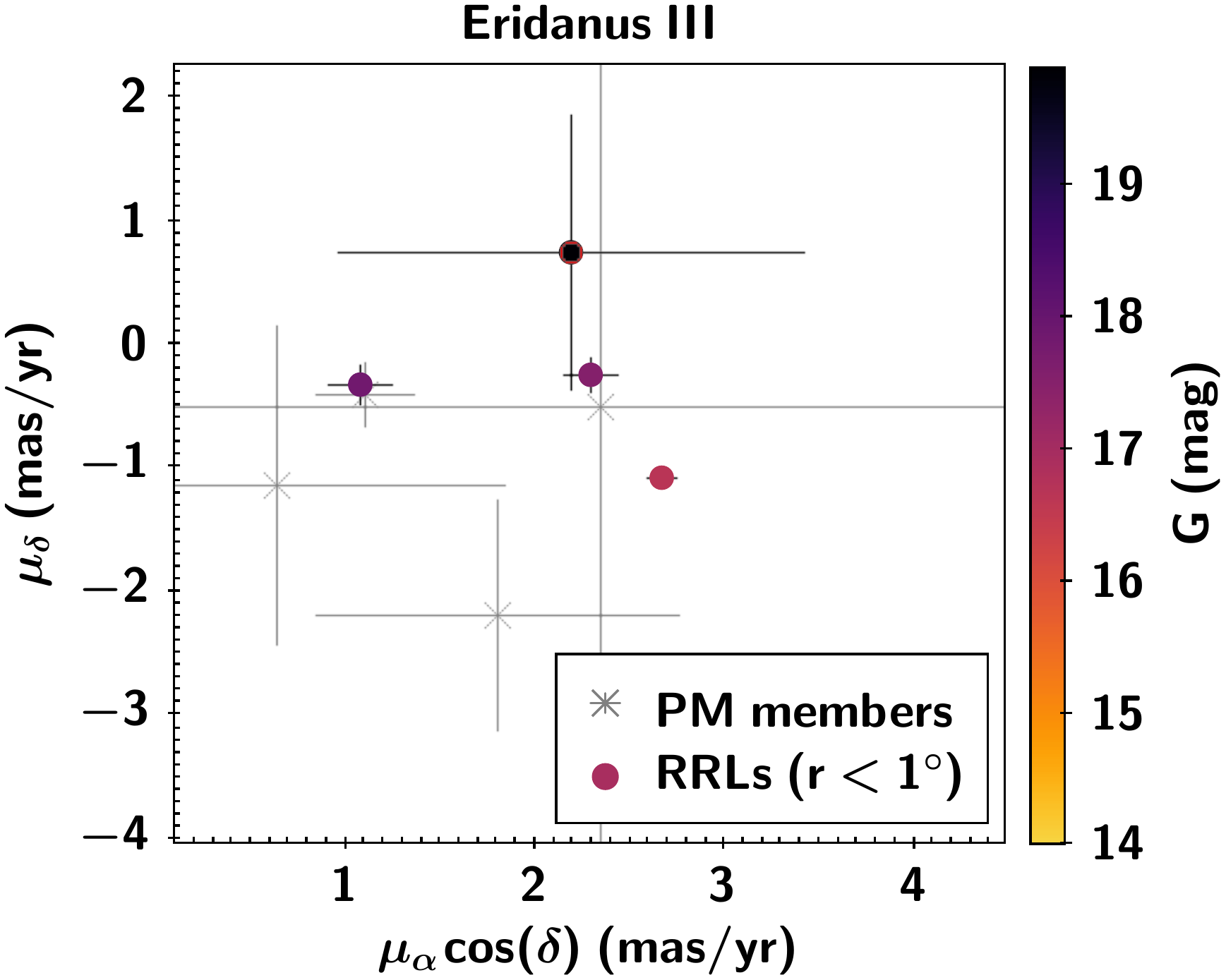}
\includegraphics[width=0.32\textwidth]{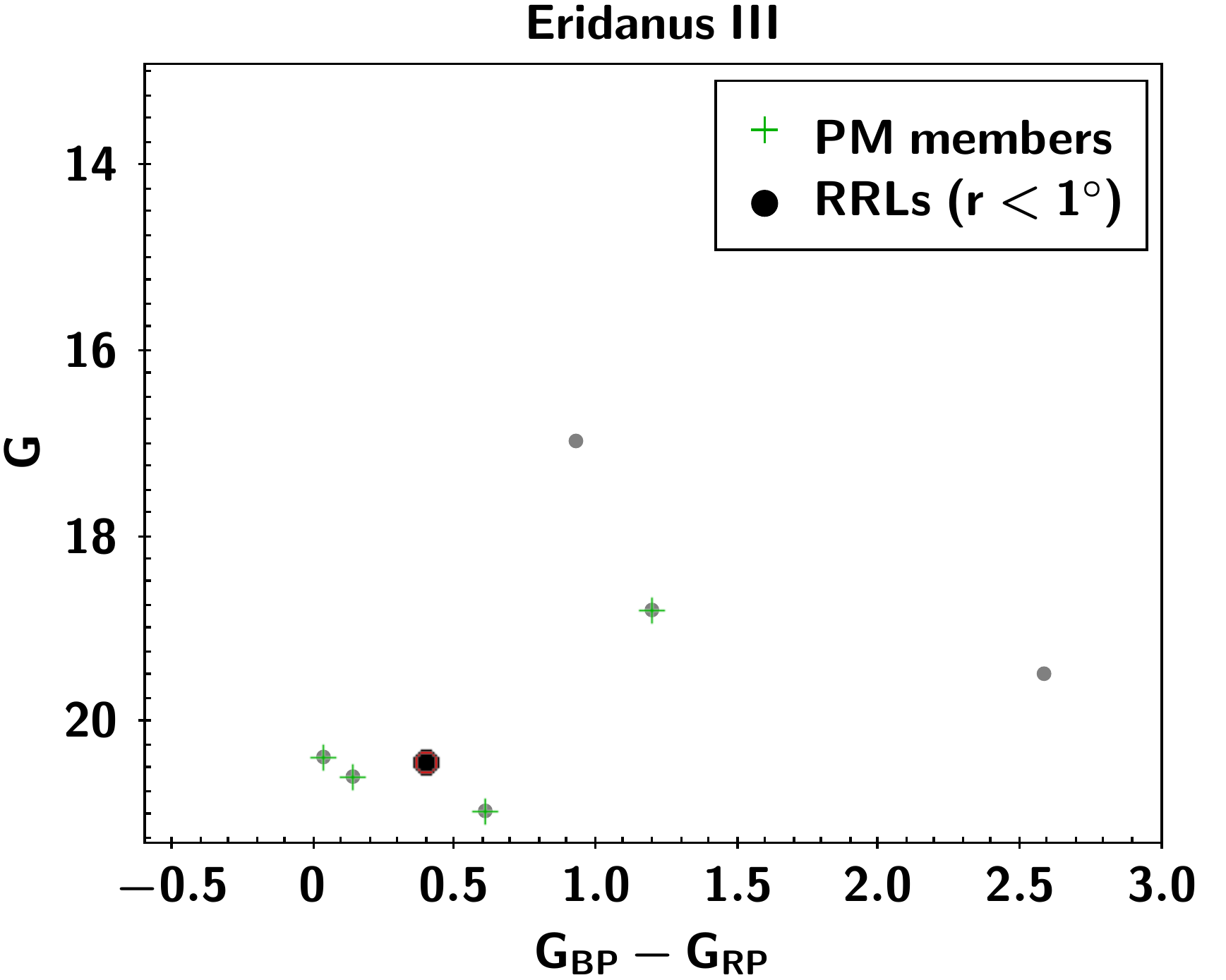}
\caption{Same as Figure~\ref{fig:UMaII} but for the Eri III UFD.}
\label{fig:EriIII}
\end{figure*}

\paragraph{Eridanus III (Figure~\ref{fig:EriIII})}

Eri III is a distant, tiny galaxy discovered in DES by \citet{bechtol15}. Its half-light radius is only $0\farcm 34$ and the tidal radius is $1\farcm 45$ \citep{munoz18}.  We found a single RRLs with the right magnitude for being a Eri III member. The star has a proper motion in agreement with other members. However, it is located at $45\arcmin$ from the center of Eri III, well outside its tidal radius. RRLs fainter than $G\sim 20.4$ are rare in the Galactic Halo \citep{medina18}, and thus the chance that this is a Galactic star is low. The lightcurve of this RRLs is noisy since it is close to the faint limit of Gaia. Photometric confirmation is desirable.

\begin{figure*}[htb!]
\centering
\includegraphics[width=0.32\textwidth]{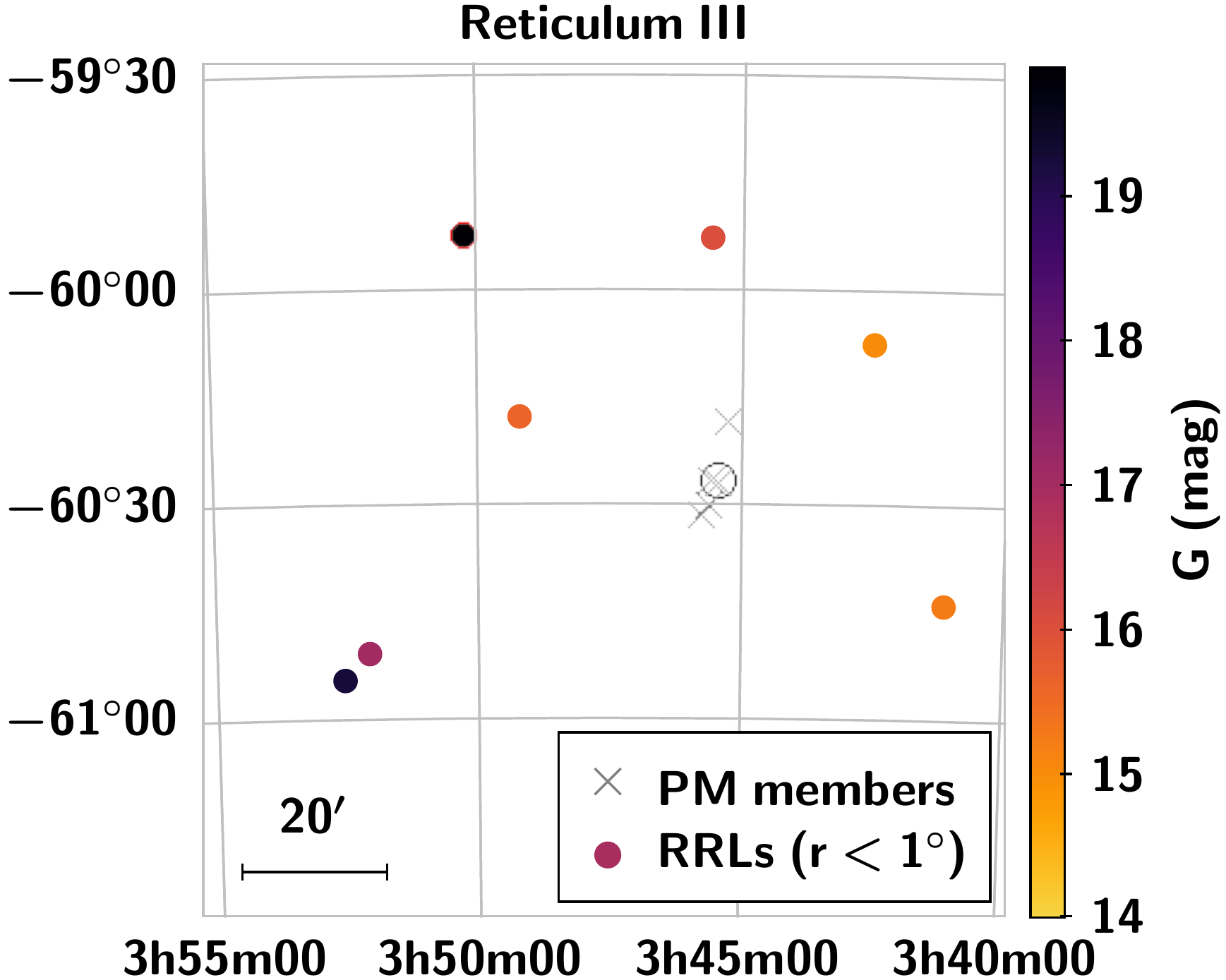}
\includegraphics[width=0.32\textwidth]{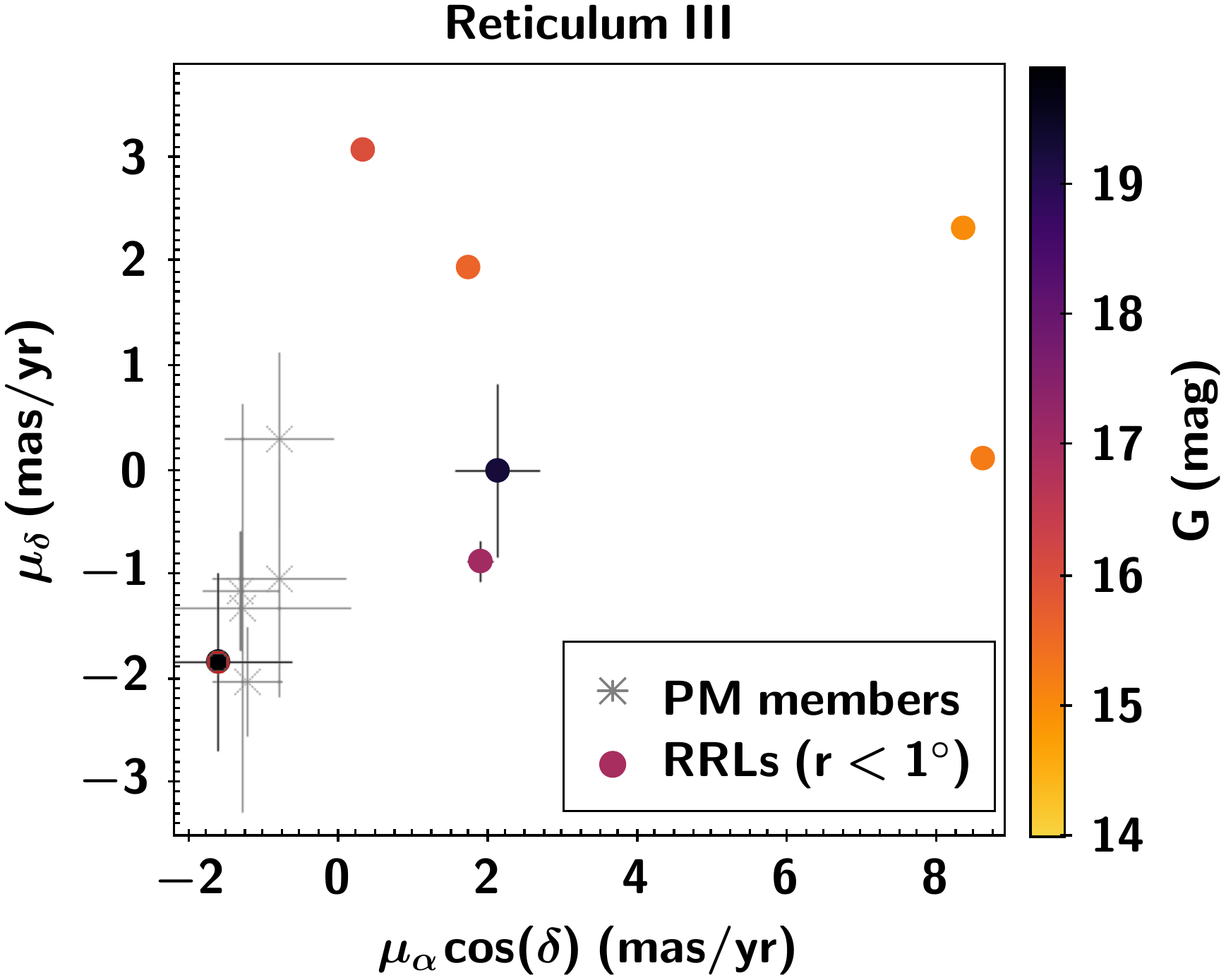}
\includegraphics[width=0.32\textwidth]{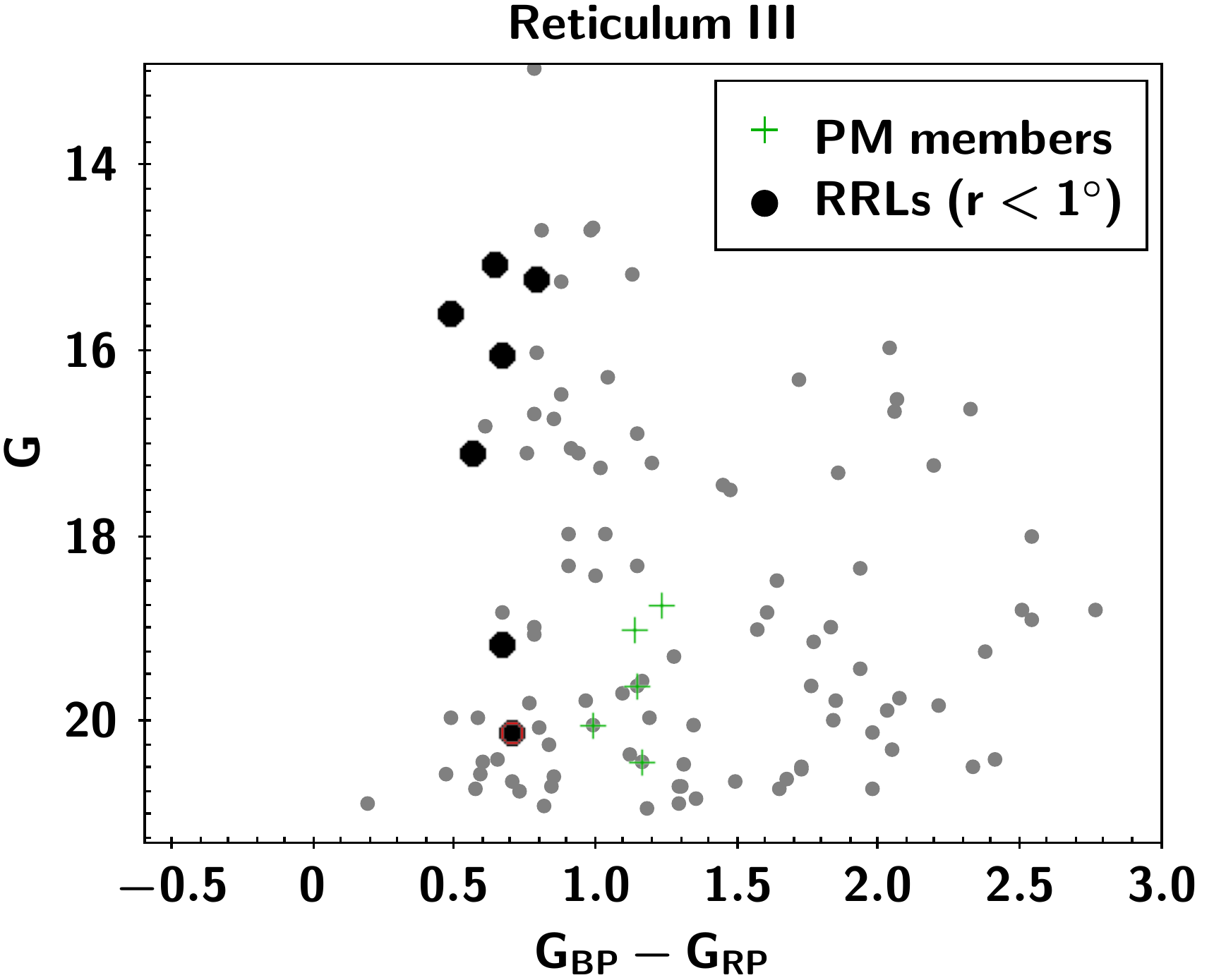}
\caption{Same as Figure~\ref{fig:UMaII} but for the Ret III UFD. Only the half-light radius ellipse is shown. }
\label{fig:RetIII}
\end{figure*}

\paragraph{Reticulum III (Figure~\ref{fig:RetIII})}

Ret III is another distant DES galaxy \citep{drlica15}. Similar to Eri III, we found also one RRLs, a type ab in this case, at $50\arcmin$ from the center of the galaxy. This star is also located well outside the tidal radius of Ret III, but being also faint, $G=20.1$, it would also be a rare chance if it is a field Halo star. The lightcurve is also noisy since this star is in the faint end for Gaia, requiring further confirmation.

\subsubsection{UFD galaxies with no new RRLs members \label{sec:nonew}}

We recovered known RRLs in four other UFD galaxies. Specifically, we recover six out of seven RRLs in UMa I, two out of three in Carina II (Car II), and one out of one in both Boo II and Phe II. The diagnostic plots for these galaxies are included as online only-material in figure set~\ref{fig:diagnostic} in the Appendix section (\S~\ref{sec:appendix}). Since Car II and Car III are very close in the sky, we show the two galaxies in the same plot. All of the recovered RRLs were confirmed as proper motion members of their respective galaxies.

\subsection{UFD Galaxies with no Gaia RRLs \label{sec:noRR}}

We did not find any Gaia RRLs associated with 13 of the UFD galaxies we explored: Cetus II (Cet II), Car III, Draco II (Dra II), Grus II, Horologium I (Hor I), Horologium II (Hor II), Reticulum II (Ret II), Segue I, Segue II, Triangulum II (Tri II), Tucana V (Tuc V), Virgo I (Vir I) and Willman I.  Figures for 12 of the 13 galaxies are included as online only material (Figure~\ref{fig:diagnostic}) in the Appendix section. No figure for Tri II is presented since no RRLs were found within our search area around that galaxy. Segue II has no Gaia RRLs either within the search area but we included a diagnostic plot in this case to confirm membership of the previously know RRLs.

Out of the 13 galaxies in this group, three of those galaxies (Segue I, Segue II, and Grus II) are known to have RRLs but Gaia did not recovered them. Based on radial velocity variations, it is suspected that Segue I contains one RRLs, although its light curve properties are still unknown \citep{simon11}. Gaia DR2 did not recover either of these stars. Although there are three RRLs within our search area, and two of them have similar magnitudes around $G =17.4$, which is also the same magnitude as the known RRL, none of those two stars agree with the proper motions of the galaxy. They are also located far away from the galaxy, at $39\arcmin$ and $49\arcmin$, which is beyond the tidal radius \citep[$16\farcm 4$,][]{munoz18}. On the other hand, Grus II and Segue II are known to have one RRL each \citep[][respectively]{martinez19,boettcher13}, but Gaia did not recover these stars either. Indeed, there are no RRLs at all within $1\degr$ of the center of Segue II. We confirm that the proper motions of the known RRLs in Grus II, Segue I and Segue II, which were taken from the main Gaia DR2 catalog,  agree with those of their respective galaxies.

Car III and Willman I have been searched before for RRLs \citep{torrealba18,siegel08} with no positive results, in agreement with our search in Gaia DR2. 

There has no been previous searches in any of Dra II, Tri II, Cet II, Ret II, Tuc V, Hor I, Hor II, and Vir I.
Although it is possible that those galaxies indeed have zero RRLs, this is far from a robust result since it is known Gaia DR2 is  quite incomplete in some parts of the sky \citep{clementini19}. This is particularly true for the last three of the above, because they are located beyond $\sim 80$ kpc and their RRLs, if they exist, would have magnitudes $G\gtrsim20.5$, close to the Gaia limiting magnitude. A careful search for variables in all these galaxies is recommended.

The case of Cet II deserves some discussion. In the field around Cet II, there are three RRLs with similar magnitudes, $G=17.4$ (Gaia IDs 2355331134227155072, 2358342868374410624 and 2358458488893773568). Although their magnitudes are close to the expectation for Cet II, their proper motions are incompatible with the ones found for this galaxy by \citet{pace19}. They are also located far from the center of Cet II, at more than $12\, r_h$. \citet{conn18} points out that the Sgr trailing arm crosses the same region of Cet II at approximately the same distance, and the wide distribution in the sky of those three RRLs suggest they may instead be Sgr stars. 

Similarly, the Chenab stream \citep{shipp18} cohabits the same region of the sky as Grus II. \citet{martinez19} found two stars which seem to be associated with the Chenab stream when looking for RRLs in that galaxy. With Gaia we recovered one of these stars (Chenab-V4), and we found another one (Gaia ID 6561477651849190912) with similar magnitude to Chenab-V4 but farther away from the center of Grus II, at $54\arcmin$. Since streams are wide structures, it is reasonable to think this star may be also a Chenab member.

In the case of Ret II we found two RRLs with very similar mean $G$ magnitudes, 18.41 and 18.46, one of them within the tidal radius of the galaxy.  Unfortunately none of them have proper motions reported in Gaia DR2. Although it was very tempting to associate them to Ret II based on their magnitude and location, the final inspection of the light curves and image stamps of those stars indicate these were actually galaxies misclassified as RRLs in Gaia. The CMD also show that these two stars are too red for being RRLs.

\section{Distances}\label{sec:distances}

We calculate the distance modulus to each RRL using the absolute magnitude, $M_G$, versus [Fe/H] relation obtained by \citet{muraveva18}:

\begin{equation}
M_G= 0.32^{+0.04}_{-0.04}\, \mathrm{[Fe/H]} + 1.11^{+0.06}_{-0.06}
\label{eq:Mg}
\end{equation}

The metallicity for each galaxy was taken from Table~\ref{tab:UFD} (column 6), and the apparent $\langle G \rangle$ measurements from Table~\ref{tab:RRL} (column 7). For Ret III, which has no available measurement of [Fe/H], we assumed the mean value for all the UFDs, which is $-2.4$ dex. For the reddening correction we used $A_G = 2.740\ E(B-V)$ \citep{casgrande18}. The excess color E($B-V$) \citep{schlegel98} was obtained using the python task \verb|dustmaps| \citep{green18}. 

The uncertainties in the distances were obtained by propagation. In this calculation, the error in the metallicity was assumed to be 0.2 dex, and the error in the extinction is conservatively considered to be the 10\% of its value. The photometric error was taken directly from the Gaia catalog (column \verb|int_average_g_error|). In addition, we included the dispersion of the $M_G$-[Fe/H] relation which turned out to be 0.14 mag after doing a Monte Carlo propagation of eq~\ref{eq:Mg}. Individual distances and their errors are shown in columns 13 and 14 of Table~\ref{tab:RRL}.

Finally, in Table~\ref{tab:distance} we show a compilation of the mean distances obtained for each system by averaging the individual distances of RRLs in each galaxy. In the averages, we included all RRLs, taking account of the extra-tidal candidates.

Distances so calculated agree with the literature values we compiled in Table~\ref{tab:UFD}, excepting four galaxies: Boo II, Eri III, and Sgr II. 

For Eri III this is the first distance estimate based on RRLs. The discovery papers of this galaxy suggest distances of 87 kpc \citep{koposov15}, and 95 kpc \citep{bechtol15}. Later, \citet{conn18b} derived a distance of $91\pm 4$ based on deep photometry of the galaxy. Our distance of $98 \pm 9$ kpc disagrees with the one given by \citet{koposov15}, but within errors agrees with those given by \citet{bechtol15} and \citet{conn18b}.

From deep photometry, \citet{longeard20} find a distance of $73.1^{+1.1}_{-0.7}$ kpc for Sgr II. The Gaia RRLs distance suggest a closer distance, $62\pm6$ kpc, which is in agreement with the distance given by \citet{joo19} of $64\pm 3$ kpc, also based in RRLs.

For the Boo II UFD, \cite{munoz18} derived a distance of 47 kpc. Instead, our distance determination using Gaia RRLs results in a closer value of $40 \pm 4$ kpc, in good agreement with the 
earlier estimate, also based on RRLs \citep[$39\pm2$ kpc,][]{sesar14}.

\vspace{1cm}
\begin{table}
\caption{Distance modulus and Heliocentric distances to the UFDs with Gaia RRLs \label{tab:distance}}
\centering
\begin{tabular}{lcccc}
\toprule
Galaxy & DM$_0$ & $\sigma_{DM_0}$ & D$_{\odot}$ & $\sigma_{D_{\odot}}$ \\
 & (mag) & (mag) & (kpc) & (kpc) \\
\hline
   Boo I & 19.04 & 0.22 & 64 &  6 \\
  Boo II & 18.00 & 0.22 & 40 &  4 \\
 Boo III & 18.34 & 0.19 & 47 &  4 \\
  Car II & 17.68 & 0.22 & 34 &  3 \\
  ComBer & 18.00 & 0.20 & 40 &  4 \\
 Eri III & 19.96 & 0.21 & 98 &  9 \\
   Hyd I & 17.31 & 0.22 & 29 &  3 \\
  Phe II & 19.99 & 0.22 & 99 & 10 \\
 Ret III & 19.70 & 0.21 & 87 &  8 \\
  Sag II & 18.97 & 0.20 & 62 &  6 \\
  Tuc II & 18.75 & 0.20 & 56 &  5 \\
 Tuc III & 17.02 & 0.21 & 26 &  2 \\
   UMa I & 19.93 & 0.19 & 97 &  9 \\
  UMa II & 17.60 & 0.20 & 33 &  3 \\
\hline
\end{tabular}
\end{table}

\begin{figure}[htb!]
\centering
\includegraphics[width=0.99\columnwidth]{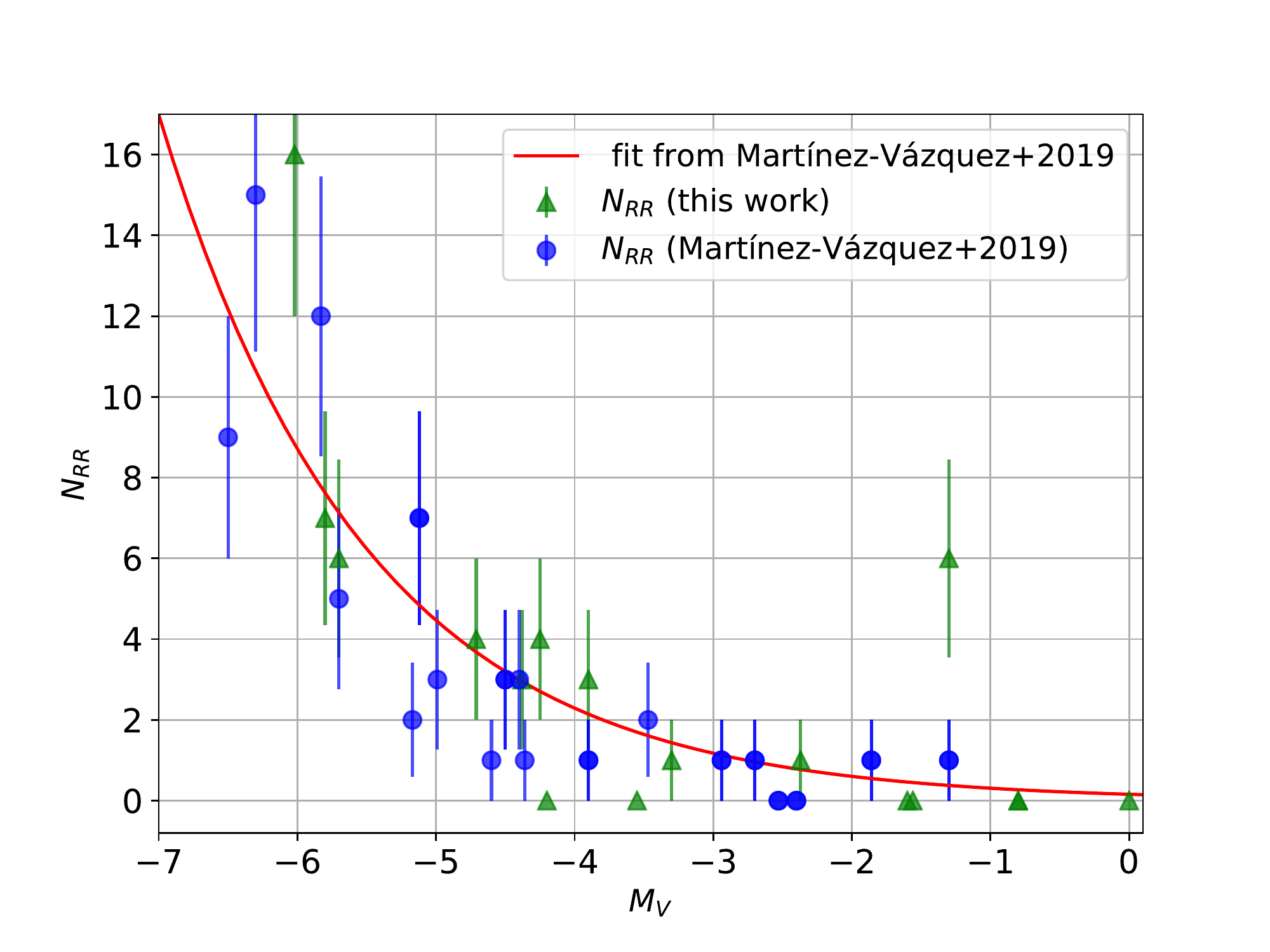}
\caption{Number of RRLs versus M$_V$ for the UFDs. Green triangles show the updated values obtained using Gaia DR2 in this work. Blue circles are the UFDs compiled by \citet{martinez19} (excluding the galaxies with updated values). Red solid line shows the fit obtained by \citet{martinez19} using all the dwarf satellites in the Milky Way and Andromeda in addition to isolated galaxies and two Sculptor group dwarfs.}
\label{fig:NRR}
\end{figure}

\begin{figure}[htb!]
\centering
\includegraphics[width=0.95\columnwidth]{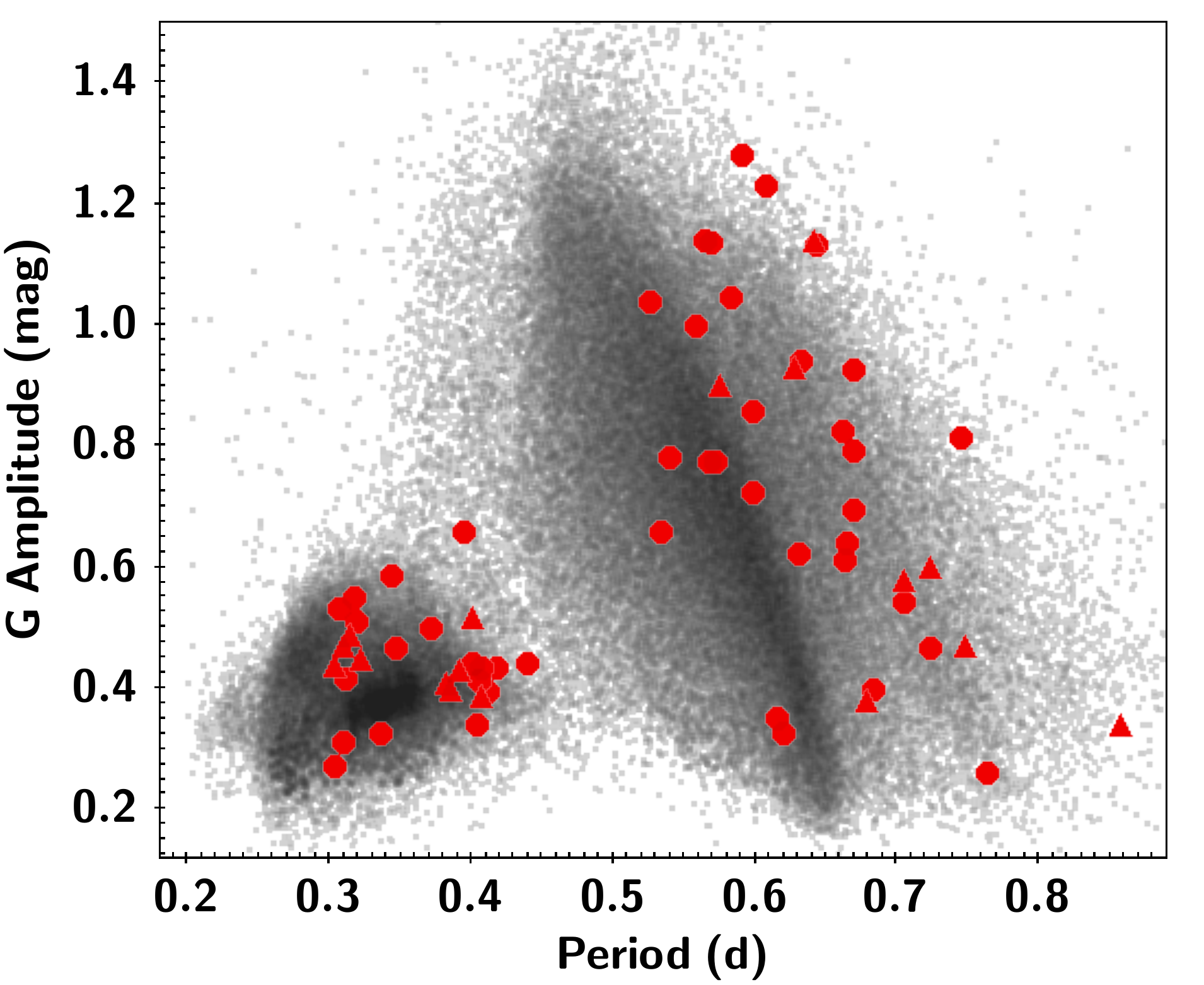}
\caption{Bailey diagram for all the RRLs in Gaia DR2 (black dots). The Gaia RRLs in the UFDs of this study are highlighted with red circles, while other RRLs not in Gaia were scaled to $G$ amplitudes and are shown as red triangles.}
\label{fig:PA}
\end{figure}

\section{Conclusions}\label{sec:conclusions}

In this work we mine the catalog of RRLs of Gaia DR2 \citep{clementini19} for such stars in 27 UFD galaxies with heliocentric distance $\le 100$ kpc, which is about the faint limit for this catalog. Although the Gaia catalog of RRLs is known to have varying completeness depending on position in the sky, we were able nonetheless to associate 47 RRLs to 14 different UFDs, including four galaxies in which no search for variable stars had been done before. About \%50 of the RRLs are being recognized for the first time as members of UFDs. We presented new distance estimates to UFDs based on the Gaia RRLs. Among the 14 UFDs with RRLs, six contain candidates to an extra-tidal population, suggesting these galaxies may be in the process of being disrupted. Further confirmation via radial velocities of those stars would be desirable. 

In Figure~\ref{fig:NRR} we show the number of known RRLs in each UFD as a function of their $M_V$. The red line is the power-law fit that \citet{martinez19} obtained using data for all dwarf satellite galaxies around the Milky Way and Andromeda in addition to isolated galaxies and two Sculptor group dwarfs, spanning near 17 magnitudes in $M_V$. The fit was done using only galaxies in which a search for RRLs has been carried out in areas enclosing $> 2\, r_h$ (i.e. the search should be complete, or close to complete). \citet{martinez19} found that several UFDs fall below that line (see their Figure 10). The updated number of RRLs obtained in this work brings most UFDs close to the fitted line. 
Tuc III, at $M_V = -1.3$ is a clear outlier in this plot. The galaxy has an unusual high number of RRLs for UFDs of similar $M_V$. Tuc III is a disrupting galaxy and, actually, all of the RRLs we found here are extra-tidal stars. The total luminosity of this galaxy may be larger if the mass lost by disruption is taken into account. This may explain the high number of RRLs found. In any case, confirmation of membership of the RRLs in this galaxy with radial velocities is needed.

The RRLs in UFDs are quite spread out in an amplitude versus period diagram (a Bailey diagram), not following a unique locus of the Oo groups. In Figure~\ref{fig:PA} we show the Bailey diagram for the Gaia RRLs in UFDs as red solid circles. We included also the RRLs not found in Gaia for the galaxies in our study, scaling their original amplitudes to the $G$ band, using the Amp $G$ - Amp $V$ relationship given in \citet{clementini16}. In the case of UMa I and Car II we first converted their $B$ and $g$ amplitudes, respectively, to $V$ scaling by 0.845 and 1.29. Those scale factors were obtained from photometry of RRLs in M68 for the $B$ band \citep{walker94}, and in Crater II for the $g$ band \citep{vivas19b}. In the background of Figure~\ref{fig:PA}  we display the location in this diagram for all 138,406 {\rrab } and {\rrc } in the Gaia catalog ({\rrd } are not plotted). The bulk of that catalog belongs to the Halo population. As expected, most of the Halo {\rrab } stars lie with the Oo I group. Thus, UFDs do not seem to be the main contributor to the Halo population of RRLs.

There are 10 UFDs within 100 kpc which seem to contain no RRLs. Only in two of them, previous dedicated searches had concluded that those galaxies indeed do not host RRLs. Because of the known Gaia completeness issues, further searches on the remaining galaxies is recommended.

\bigskip

\acknowledgments

We are indebted to the anonymous referee for a constructive review of the paper.
This work has made use of data from the European Space Agency (ESA) mission
{\it Gaia} (\url{https://www.cosmos.esa.int/gaia}), processed by the {\it Gaia}
Data Processing and Analysis Consortium (DPAC,
\url{https://www.cosmos.esa.int/web/gaia/dpac/consortium}). Funding for the DPAC
has been provided by national institutions, in particular the institutions
participating in the {\it Gaia} Multilateral Agreement.

\facility{Gaia}

\software{Topcat v4.7 \citep{topcat}, Aladin \citep{aladin1, aladin2}, Matplotlib \citep{matplotlib}}

\appendix

\section{Online-only material \label{sec:appendix}}

\begin{figure}
\figurenum{13}
\epsscale{1.15}
\plotone{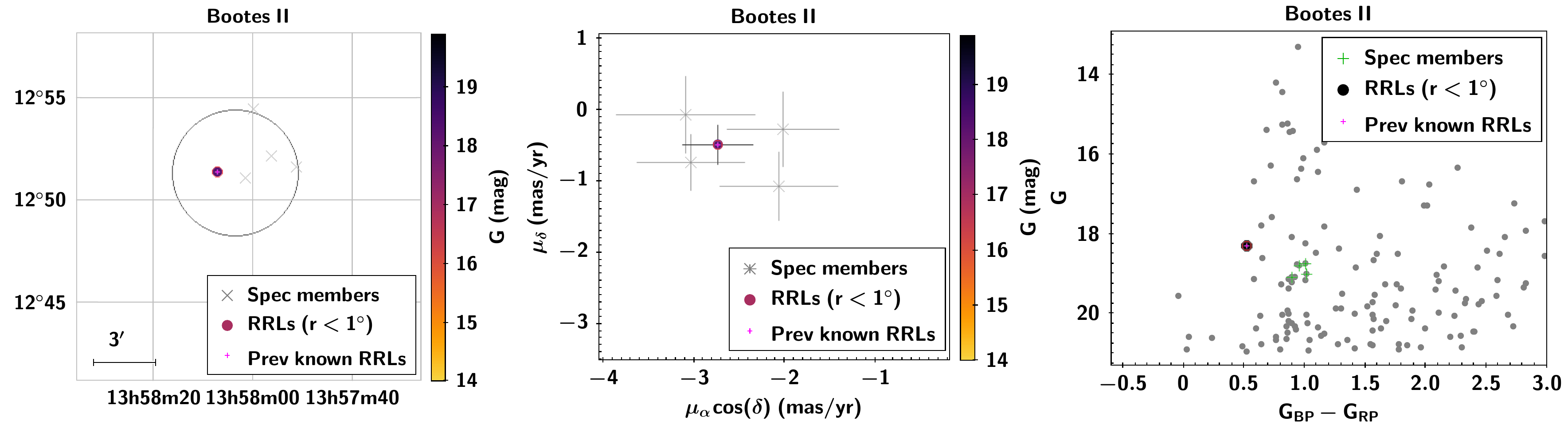}
\caption{The left and middle panels show the map in equatorial coordinates and the proper motions of RRLs within a $1\degr$ radius of the galaxy (solid circles). 
In both panels, the color of the circles scale with the mean $G$ magnitude of the star, and spectroscopic (or proper motion) members are shown as grey crosses. 
Previously known RRLs in this galaxy are marked with magenta $+$ symbols.  In the left panel, the ellipse indicates the half-light radius (see Table~\ref{tab:UFD}). 
The right panel is a Gaia CMD (grey background) of stars within $1\, r_h$ from its center. 
If there are RRLs which are members of the galaxy they are enclosed by open red circles in all panels.}
\label{fig:diagnostic}
\end{figure}

\begin{figure}
\figurenum{14}
\epsscale{0.6}
\plotone{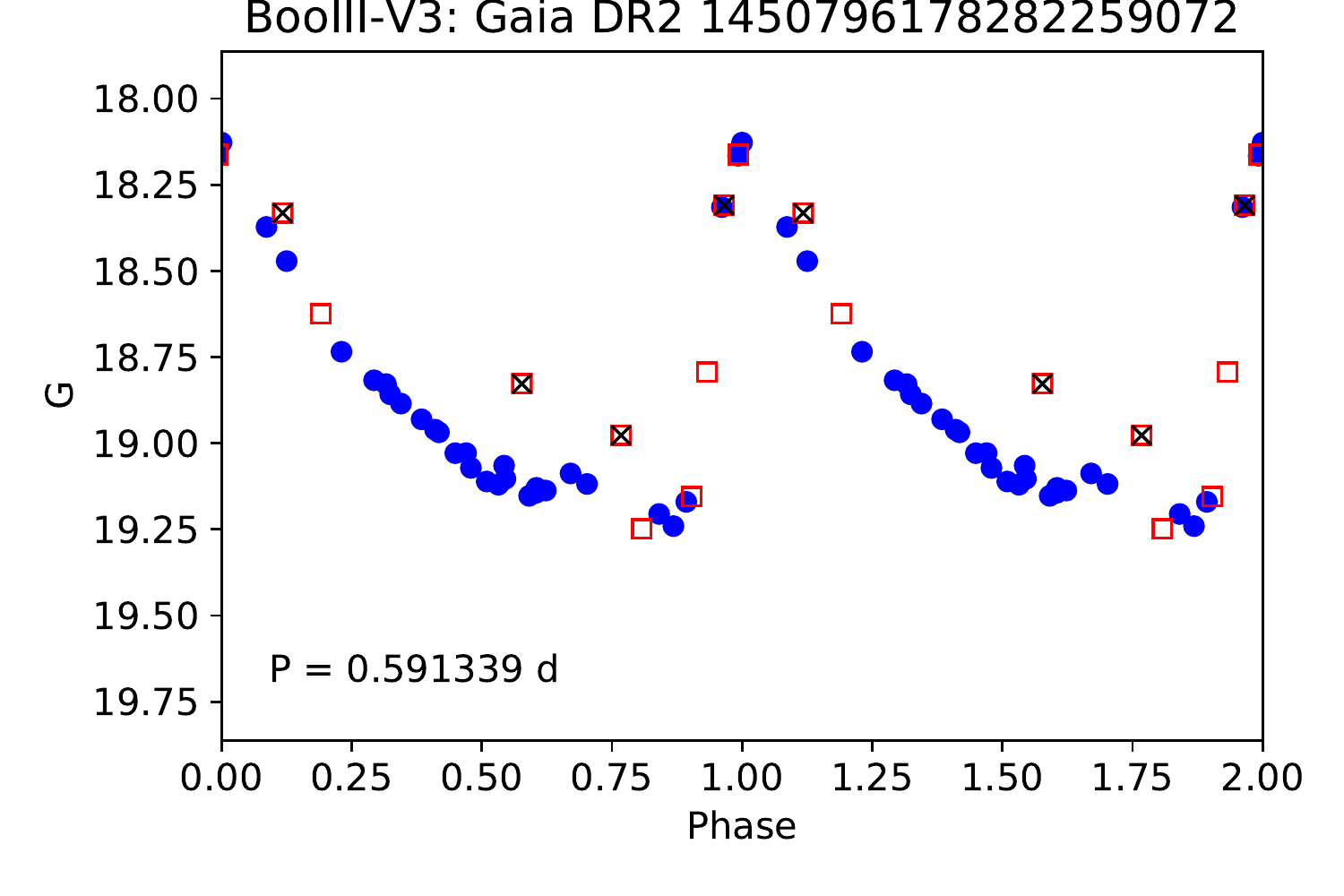}
\caption{Gaia G magnitude phased lightcurve. Red squares and black $x$'s mark measurements flagged as noisy and rejected, respectively.}
\label{fig:lc}
\end{figure}

We provide here two Figure sets available in the online version of the Journal. The first set (Figure~\ref{fig:diagnostic}) contains the diagnostic plots for all galaxies with no new RRLs to report (\S~\ref{sec:nonew}) or with no RRLs at all (\S~\ref{sec:noRR}). The second set (Figuire~\ref{fig:lc}) are the $G$-band (from Gaia) phased lightcurves for all RRLs in Table~\ref{tab:RRL}.

\figsetstart
\figsetnum{13}
\figsettitle{Diagnostic plots for galaxies with no new Gaia RR Lyrae or no RR Lyrae stars at all as described in Sections~\ref{sec:nonew} and \ref{sec:noRR}}

\figsetgrpstart
\figsetgrpnum{13.1}
\figsetgrptitle{Diagnostic plots for Bootes II}
\figsetplot{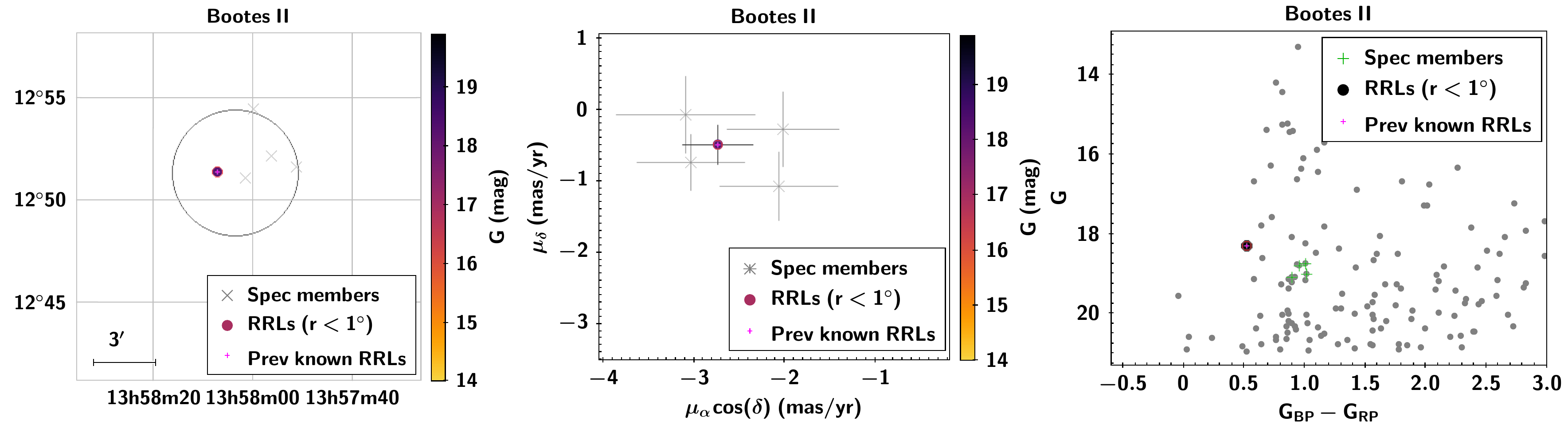}
\figsetgrpnote{The left and middle panels show the map in equatorial coordinates and the proper motions of RRLs within a $1\degr$ radius of the galaxy (solid circles). 
In both panels, the color of the circles scale with the mean $G$ magnitude of the star, and spectroscopic (or proper motion) members are shown as grey crosses. 
Previously known RRLs in this galaxy are marked with magenta $+$ symbols.  In the left panel, the ellipse indicates the half-light radius (see Table~\ref{tab:UFD}). 
The right panel is a Gaia CMD (grey background) of stars within $1\, r_h$ from its center. 
If there are RRLs which are members of the galaxy they are enclosed by open red circles in all panels.}
\figsetgrpend

\figsetend

\figsetstart
\figsetnum{14}
\figsettitle{Gaia G-magnitude phased light curves for RR Lyrae stars in Table 2.}

\figsetgrpstart
\figsetgrpnum{14.7}
\figsetgrptitle{Phased lightcurve for star BooIII-V3}
\figsetplot{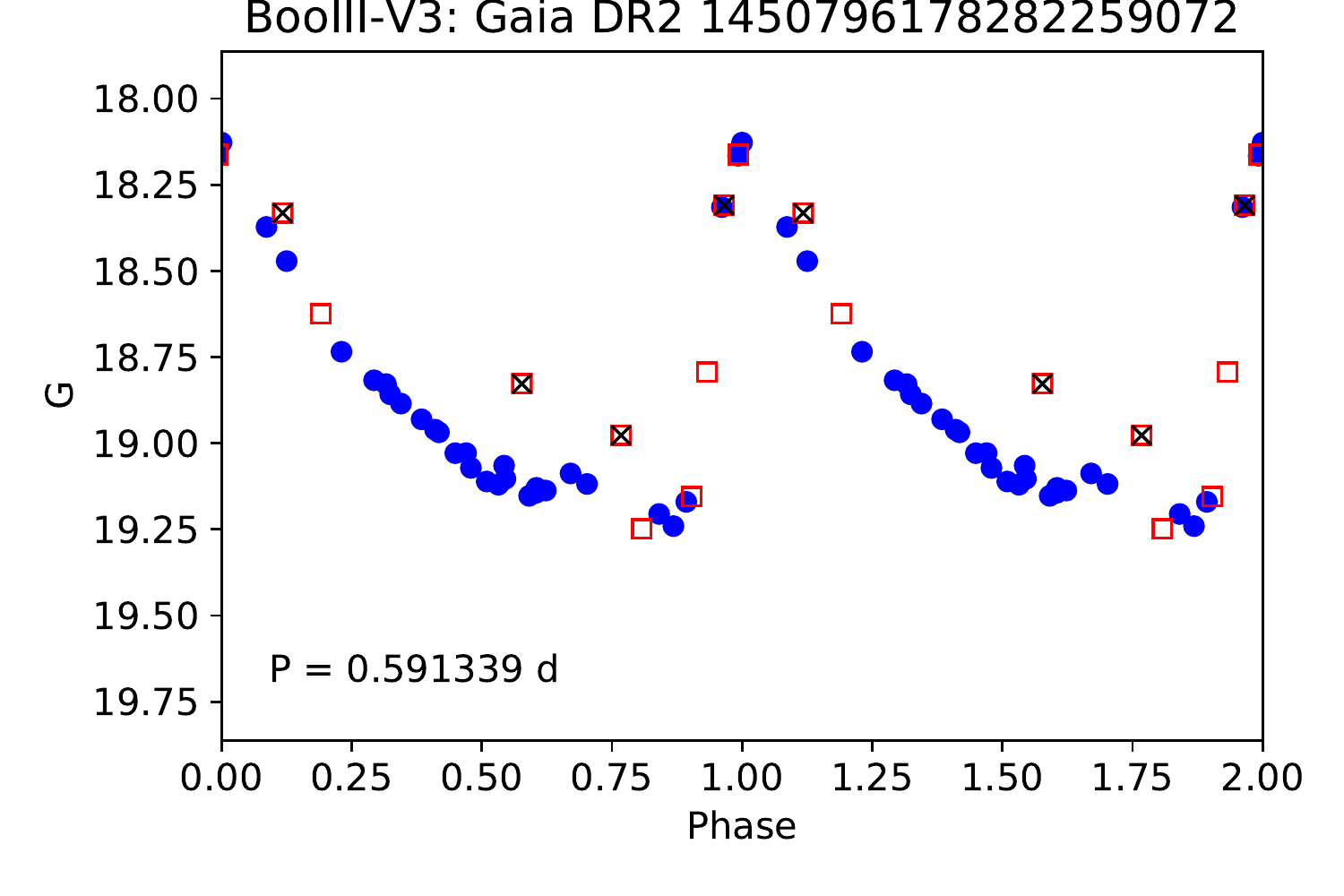}
\figsetgrpnote{Gaia G magnitude phased lightcurve. Red squares and black $x$'s mark measurements flagged as noisy and rejected, respectively.}
\figsetgrpend

\figsetend

%%%%%%%%%%%%%%%%%%%%%%%%%%%%%%%%%%%%%%%%%%%%%%%%%%

%\bibliographystyle{aasjournal}
%\bibliography{KV}

%%%%%%%%%%%%%%%%%%%%%%%%%%%%%%%%%%%%%%%%%%%%%%%%%%

\end{document}